\documentclass[12pt]{article}
\usepackage{amsmath}
\usepackage{amssymb}
\usepackage{epsfig}
\newcommand{\be}{\begin{equation}}
\newcommand{\ee}{\end{equation}}
\newcommand{\bea}{\begin{eqnarray}\displaystyle}
\newcommand{\eea}{\end{eqnarray}}
\newcommand{\bdm}{\begin{displaymath}}
\newcommand{\edm}{\end{displaymath}}

\newcommand{\Tr}{\mathop{\rm Tr}\nolimits}

\def\ket#1{|#1 \rangle}

\def\aver#1{\langle\, #1 \,\rangle}

\def \ll {{\cal L}}

\def \Bhat {{\widehat{\cal B}}}
\def \Uhat {{\widehat{U}}}

\def \lbra{\Tr(}
\def \Llbra{\Tr\left(}
\def \Blbra{\Tr\Big(}

\def \rbra{)}
\def \Rrbra{\right)}
\def \Brbra{\Big)}

\numberwithin{equation}{section}
\numberwithin{figure}{section}

\begin{document}

\def\half{\fraction{1}{2}}
\def\fraction#1#2{ { \scriptstyle \frac{#1}{#2} }}
\def\d{\partial}
\def\lineup{\!\!\!\!\!\!\!\!\!&&}
\def\EQ#1{eq.(\ref{eq:#1})}

\def\B{\mathcal{B}_0}
\def\L{\mathcal{L}_0}

\begin{titlepage}
\rightline{\today}
{}~\hfill\vbox{\hbox{hep-th/yymmnnn}\hbox{NSF-KITP-09-23}
}\break \vskip 2.1cm

\begin{center}
\vskip 1.0cm {\large \bf{A Simple Analytic Solution for Tachyon Condensation} }
\\
\vskip 1.0cm

{\large Theodore Erler$^{a,b,}$\footnote{Email: tchovi@gmail.com},
Martin Schnabl$^{a,b,}$\footnote{Email: schnabl.martin@gmail.com}}
\vskip 1.0cm

$^a${\it {Institute of Physics of the ASCR, v.v.i.} \\
{Na Slovance 2, 182 21 Prague 8, Czech Republic}}
\vskip .5cm
$^b${\it {Kavli Institute for Theoretical Physics} \\
{University of California, Santa Barbara, CA 93106-4030, USA} }

\vskip 1.0cm
{\bf Abstract}
\end{center}

\noindent In this paper we present a new and simple analytic solution for
tachyon condensation in open bosonic string field theory. Unlike the
$\mathcal{B}_0$ gauge solution, which requires a carefully regulated discrete
sum of wedge states subtracted against a mysterious ``phantom'' counter term,
this new solution involves a continuous integral of wedge states, and no
regularization or phantom term is necessary. Moreover, we can evaluate the
action and prove Sen's conjecture in a mere few lines of calculation.

\medskip

\end{titlepage}

\newpage

\baselineskip=18pt

\tableofcontents

\section{Introduction} The original analytic solution for tachyon
condensation in open bosonic string field theory \cite{Schnabl}
(henceforth, the $\mathcal{B}_0$ gauge solution) takes the
form of a regulated sum
\begin{equation}\Phi = \lim_{N\to\infty}\left[\psi_N -
\sum_{n=0}^N\frac{d}{dn}\psi_n
\right],\end{equation}
where $\psi_n$ are wedge states with certain insertions (for more details, see
\cite{Schnabl,Okawa}). The form of this solution has long been a
puzzle. First, the limit suggests that the solution may live outside the space
of well-behaved string fields---like a distribution is a limit of a sequence
of functions. Second, the mysterious $\psi_N$ term---the so-called
``phantom piece''---actually {\it vanishes} when contracted with well-behaved
states in the large $N$ limit. But we cannot simply set
$\lim_{N\to\infty}\psi_N=0$ since, if we evaluate the action
analytically \cite{Schnabl}, the $\psi_N$ term produces a substantial portion
of the energy required to prove Sen's conjecture \cite{Sen}. Yet, the $\psi_N$
term does not contribute to the energy in the ordinary level
expansion \cite{Schnabl,Takahashi}, since as a state in the Fock space it
vanishes identically.

By now the regularization and phantom piece are better
understood \cite{Okawa,Fuchs,Aldo,Erler2,Erler3,Russians,Russians2},
and there is little
doubt that the $\mathcal{B}_0$ gauge solution is for practical purposes
nonsingular. Yet,
no one has found an adequate definition of the solution---or gauge equivalent
alternative---which does not require the regulated sum and phantom piece.

In this note, we present an alternative solution for the tachyon vacuum which
avoids the above complications. Instead of a discrete sum, the solution
involves a continuous integral over wedge states, and no regularization or
mysterious phantom term is necessary. Moreover, evaluation of the action and
the proof of Sen's conjectures is, in contrast to the $\mathcal{B}_0$ gauge,
very straightforward.

Broad classes of generalizations of the $\B$ gauge solution have been
constructed in \cite{RZ,RZO,K&M,Erler1,Erler2}. Note in particular that
our new solution is a special case of the solutions considered in
\cite{Erler2}, though our analysis will be quite different.

This paper is organized as follows. After some algebraic and notational
preliminaries, in Section \ref{sec:solution} we present the new solution for
the tachyon vacuum, comment on its structure, and prove the equations of
motion. In Sec.\ref{subsec:Sen} we prove
Sen's conjectures, specifically proving the absence of open string states
and giving a very simple calculation of the brane tension. In
Sec.\ref{subsec:gauge} we comment on the relation between pure gauge 
solutions and the phantom piece, and in Sec.\ref{subsec:Ian} we compute the 
closed string tadpole and demonstrate that it vanishes. In Section 
\ref{sec:level} we investigate the energy of the new vacuum
in level truncation. As a warmup exercise, in Sec.\ref{subsec:Curly} we
consider the $\mathcal{L}_0$ level expansion. Due to the remarkable simplicity
of our solution, we can solve the $\mathcal{L}_0$ expansion exactly; we
resum the expansion to confirm Sen's conjecture up to better than one part
in 10 million. In Sec.\ref{subsec:Square} we consider the ``true'' level
expansion in terms of eigenstates of $L_0$. Surprisingly---unlike the Siegel
gauge or $\B$ gauge tachyon condensates---we find that the expansion for
the energy does not converge. In order to understand this phenomenon, in
section Sec.\ref{subsec:Id_corr} we consider a toy model of our
solution where the $L_0$ level expansion, though divergent, can be
solved exactly. In the end, we are able to resum the $L_0$ expansion of our 
solution and confirm Sen's conjecture to better than $99\%$. We end with some 
discussion.

\section{Solution}
\label{sec:solution}

The new vacuum solution can be presented using the same basic algebraic setup
as the original $\mathcal{B}_0$ gauge solution \cite{Okawa,Erler1}---that is,
it can be built out of three ``atomic'' string fields $K,B,c$:
\begin{eqnarray}K = \lineup \mathrm{Grassmann\ even,\ gh}\#=0,\nonumber\\
B = \lineup \mathrm{Grassmann\ odd,\ gh}\#=-1,\nonumber\\
c = \lineup \mathrm{Grassmann\ odd,\ gh}\#=1 ,
\end{eqnarray}
which satisfy the algebraic relations
\begin{eqnarray}[K,B] = 0,\ \ \ \lineup Bc+cB=1,\nonumber\\
B^2=0,\ \ \ \lineup c^2=0,\label{eq:alg_id}\end{eqnarray} and have BRST
variations ($Q=Q_B$)
\begin{equation}QK =0,\ \ \ QB=K,\ \ \ Qc = cKc.\label{eq:BRST_id}
\end{equation}
All products above are open string star products. Thus, $K,B,c$ generate a
subalgebra of the open string star algebra which is closed under the action of
the BRST operator. Perhaps the most useful explicit definition of $K,B,c$ is
given in terms of CFT correlation functions on the
cylinder\footnote{In the operator notation these fields can be written,
\begin{equation} K = \frac{\pi}{2} (K_1)_L
\ket{I},\ \ \  B = \frac{\pi}{2} (B_1)_L \ket{I},\ \ \ c = \frac{1}{\pi}c(1)
\ket{I},\end{equation} where $K_1=L_1+L_{-1},B_1=b_1+b_{-1}$, $\ket{I}$ is the
identity string field, and the subscript
$L$ denotes taking the left half of the corresponding charge---that is,
integrating the current from $-i$ to $i$ on the positive half of the unit
semicircle. Note that each field $K,B,c$ written here differs by a sign from
the definitions used in \cite{Erler1,Erler2}.}. To keep the presentation
self-contained, we explain how this works in appendix \ref{app:cylinder}.
Note that the $SL(2,\mathbb{R})$ vacuum can be written explicitly in terms of
$K$ \cite{Okawa,Erler1}:
\begin{equation}|0\rangle \equiv \Omega = e^{-K}. \end{equation}
By extension, any power of the vacuum---that is, a wedge state
\cite{RZ_wedge}---can be expressed as $\Omega^t = e^{-tK}$ for
$t\geq 0$.

\bigskip

\begin{figure}[t]
\begin{center}
\resizebox{5.0in}{1.6in}{\includegraphics{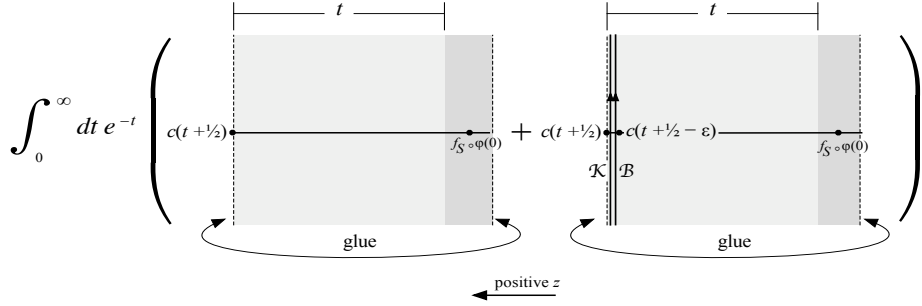}}
\end{center}
\caption{\label{fig:solution} Overlap of the solution \EQ{solution} with a Fock
space state $|\phi\rangle$, pictured as a conformal field theory correlation
function on the cylinder. See appendix \ref{app:cylinder} for further
explanation.}\end{figure}

With these preparations, the new solution for the tachyon vacuum is:
\begin{equation}\Psi = \Big[c+cKBc\Big]\frac{1}{1+K}.
\label{eq:solution}\end{equation} Let us be specific about the definition of
$\frac{1}{1+K}$. We can invert $1+K$ using the Schwinger parameterization
\begin{equation}\frac{1}{1+K} = \int_0^\infty dt\, e^{-t(1+K)} =
\int_0^\infty dt\, e^{-t}\Omega^t,\label{eq:1mKinv}\end{equation}
so, if we like, we can re-express \EQ{solution} in the form
\begin{equation}\Psi = \int_0^\infty dt\, e^{-t}\Big[c+cKBc\Big]\Omega^t.
\label{eq:solution2}\end{equation}
That's all there is to it. No regularization or ``phantom piece'' is
necessary. See figure \ref{fig:solution} for a picture of the solution as
a correlation function on the cylinder.

It is straightforward to verify the equations of motion. Note that
$cKBc=Q(Bc)$ and hence
\be\label{QPsi}
Q\Psi = cKc\frac{1}{1+K}.
\ee
To compute $\Psi^2$ it is convenient to write $c+cKBc$ as $c(1+K)Bc$.
Then commute one of the $B$s in $\Psi^2$ towards the other and the
equations of motion are quickly established.

An important property of our solution is that it involves integration over
wedge states arbitrarily close to the identity. The identity
string field is a somewhat unruly object \cite{RZ_wedge,Schnabl_wedge}, and
indeed the solution exhibits surprising convergence properties in
the level expansion. But still we have found convincing analytic and
numerical evidence that the solution describes the endpoint of tachyon
condensation. We explicitly construct the gauge transformation
relating this solution to the $\B$ gauge vacuum in appendix \ref{app:bgauge}.

Eq.(\ref{eq:solution}) is closely related to another solution which satisfies
the string field reality condition\footnote{In open string field theory, the
string field is conventionally assumed to satisfy the following reality
condition: $$\Phi^\ddag=\Phi,$$ where $\ddag$ is an involution of the star
algebra defined by the composition of BPZ and Hermitian conjugation
\cite{Zwiebach}. $K,B$ and $c$ are real string
fields in this sense, so in this context the reality condition simply
requires that the string field read the same way from the left as from the
right. The reality condition is sufficient to guarantee that the action is
real and that the string field carries the correct number of perturbative
degrees of freedom. However, all known observables in string field theory
are invariant under ``complex'' gauge transformations which do not
necessarily preserve the reality condition. Therefore an acceptable solution
may not satisfy the reality condition, but it must be in the same (complex)
gauge orbit as a solution that does.}:
\begin{equation}\hat{\Psi} = \frac{1}{\sqrt{1+K}}\Big[c+cKBc\Big]
\frac{1}{\sqrt{1+K}},\label{eq:real_sol}\end{equation}
where the inverse square root of $1+K$ is
\begin{equation}\frac{1}{\sqrt{1+K}} = \frac{1}{\sqrt{\pi}}
\int_0^\infty dt\, \frac{1}{\sqrt{t}} e^{-t}\Omega^t.\end{equation}
$\Psi$ and $\hat{\Psi}$ are related by a complex homogeneous gauge
transformation
\begin{equation}\hat{\Psi} = \frac{1}{\sqrt{1+K}}(Q+\Psi)\sqrt{1+K}.
\end{equation}
The original $\Psi$ is a simpler solution, but for some purposes the real
$\hat{\Psi}$ is more convenient. For example, $\hat{\Psi}$ is twist
even, so it lives in the same universal subspace as the $\B$ gauge vacuum
and the Siegel gauge condensate. Also, the non-real $\Psi$ has a $c$
insertion on the boundary of the local coordinate, so $\Psi$ could have
singular contractions with states carrying insertions that collide with
the $c$ ghost\footnote{Note that this problem may also afflict $\hat{\Psi}$;
though the $c$ insertion never sits on the boundary of the local
coordinate, it becomes arbitrarily close to the boundary as the integration
approaches the identity string field. Hence, for example, the action
of the operators $b(1)$ and $b(-1)$ on both $\Psi$ and $\hat{\Psi}$ is 
divergent due to singular collisions with the $c$-ghost.}.
For the purposes of this paper these differences will not prove to be
significant. The analytic proof of Sen's conjectures is identical for either
solution, and we will often use them interchangeably.

Neither $\Psi$ nor $\hat{\Psi}$ satisfies a linear $b$-ghost gauge condition.
However they do satisfy a linear gauge of a more general type,
something we call a ``dressed $\B$ gauge.'' We will explain this
class of gauges in appendix \ref{app:gauge_fix}.

\subsection{Sen's Conjectures}
\label{subsec:Sen}

Let us demonstrate that the solution (\ref{eq:solution}) describes the 
endpoint of tachyon condensation. We need to establish two things \cite{Sen}: 
first, no open strings are present at the vacuum, and second, that the vacuum 
has precisely minus the energy of an unstable D-brane.

It is easy to show that $\Psi$ supports no open string excitations. Following
\cite{cohomology1,cohomology2}, this follows if there exists a string
field $A$ (the homotopy operator) satisfying
\begin{equation}Q_\Psi A =1,\end{equation}
where $Q_\Psi=Q+[\Psi,\cdot]$ is the vacuum kinetic operator. If this is the
case, any $Q_\Psi$ closed state $\Phi$ can be written as $Q_\Psi(A\Phi)$ and
the cohomology is trivial. The homotopy operator for our solution is easily
found:
\begin{equation}A= B\frac{1}{1+K}.\label{eq:hom}\end{equation}
Therefore $Q_\Psi$ has no cohomology\footnote{We should mention that the
existence of a homotopy operator implies the absence of cohomology at all
ghost numbers, not just at the physical ghost number of $1$. This appears
to be in conflict with some numerical studies \cite{Imbimbo}, and the paradox
has yet to be resolved.}.

Let us now calculate the energy. Sen's conjecture predicts that, in the
appropriate units\footnote{We normalize the ghost correlator
\begin{equation}\langle c(z_1)c(z_2)c(z_3)\rangle_{\mathrm{UHP}}
= (z_1-z_2)(z_2-z_3)(z_1-z_3)\end{equation}
and set the spacetime volume factor and open string coupling constant to
unity. Our normalizations agree with \cite{Schnabl,Okawa}.},
the energy of the vacuum should be
\begin{equation}E = -S(\Psi) = -\frac{1}{2\pi^2},\end{equation}
where $S(\Psi)$ is the action. Assuming the equations of motion, we can
compute the action using only the kinetic term:
\begin{equation} E = \frac{1}{6}\langle \Psi, Q_B\Psi\rangle
=\frac{1}{6}\Llbra\Big[ c+cKBc]\frac{1}{1+K}cKc\frac{1}{1+K}\Rrbra,
\end{equation}
where we write
\begin{equation}\lbra \cdot \rbra= \langle I,\cdot\rangle \end{equation}
to denote the one point vertex. Now expand the
$\frac{1}{1+K}$ factors in terms of wedge states and use
$cKBc=Q(Bc)$ to write the second term as a ``total derivative'':
\begin{equation}E = \frac{1}{6}\int_0^\infty dt_1dt_2\,e^{-t_1-t_2}
\left[\Blbra c\Omega^{t_1}cKc\Omega^{t_2}\Brbra -
\Llbra Q\Big[Bc\Omega^{t_1}cKc\Omega^{t_2}\Big]\Rrbra\right].
\end{equation}
The second term is a trace of a BRST exact state, and therefore
vanishes\footnote{One should be a little careful about this. In particular,
since the integration includes traces of wedge states arbitrarily close to
the identity, if the insertions have net scaling dimension $\geq 2$ in the
sliver coordinate frame, there could be a divergence leading to an anomaly.
Fortunately, the insertions in the second term have net scaling dimension
$-1$, so such divergences are absent.}. The energy reduces to:
\begin{equation}E = \frac{1}{6}\int_0^\infty dt_1dt_2\, e^{-t_1-t_2}
\Blbra c\Omega^{t_1}cKc\Omega^{t_2}\Brbra.\end{equation}
Following appendix \ref{app:cylinder}, we can translate the trace into a
correlation function on the cylinder, which is then easy to evaluate by the
usual CFT methods. (This particular correlator has already been computed
e.g. in \cite{Schnabl,Okawa}.) The answer is,
\begin{equation}\Blbra c\Omega^{t_1}cKc\Omega^{t_2}\Brbra =
-\left(\frac{t_1+t_2}{\pi}\right)^2 \sin^2 \frac{\pi t_1}{t_1+t_2}.
\end{equation}
Therefore, we can compute the energy by evaluating the double integral,
\begin{equation}E = -\frac{1}{6}\int_0^\infty dt_1dt_2\, e^{-t_1-t_2}
\left(\frac{t_1+t_2}{\pi}\right)^2 \sin^2 \frac{\pi t_1}{t_1+t_2}.
\end{equation}
This looks complicated, but with the substitution
\begin{eqnarray}u = \lineup t_1+t_2, \ \ \ u\in[0,\infty),\nonumber\\
v = \lineup\frac{t_1}{t_1+t_2},\ \ \ v\in[0,1],\nonumber\\
dt_1dt_2= \lineup u\ \! dudv,\end{eqnarray}
the double integral factorizes into a product of two very simple integrals
\begin{equation}E=-\frac{1}{6\pi^2}\left(\int_0^\infty du\, u^3 e^{-u}\right)
\left(\int_0^1 dv \sin^2 \pi v\right).
\end{equation}
The first is $\Gamma(4) = 6$, and the second is the integral
of $\sin^2$ over a period, which produces a factor of $1/2$. Therefore
\begin{equation}E= -\frac{1}{2\pi^2}\end{equation}
in agreement with Sen's conjecture.

\subsection{Pure Gauge Solutions and the Phantom Piece}
\label{subsec:gauge}

The absence of a phantom term in our solution comes as a surprise. To see why,
let us mention a related issue: All solutions for the tachyon vacuum
(constructed so far) are, in a sense, arbitrarily close to being pure gauge.
In particular, for every vacuum solution $\Phi$, there is a one parameter
family of pure gauge solutions $\Phi_\lambda, \lambda \in[0,1)$ such that the
Fock space component fields of $\Phi_\lambda$ approach those of $\Phi$ as
$\lambda$ approaches 1. Yet, if the tachyon vacuum is expanded in a basis
of $\mathcal{L}_0$ eigenstates (see next section) the expansion coefficients
never appear close to a pure gauge solution, for any $\lambda$. Therefore the
tachyon vacuum and pure gauge solutions must differ by a term which vanishes
in the Fock space, but whose expansion in $\mathcal{L}_0$ eigenstates is
nevertheless nonvanishing. This is the origin of the phantom piece.

Since the phantom piece does not explicitly appear in our solution, we need
to track down where it went. Following Okawa \cite{Okawa}\footnote{The Okawa
pure gauge form for our solution is
\begin{equation}\Psi_\lambda=(1-\lambda\Phi)Q\frac{1}{1-\lambda\Phi},\ \ \
\Phi=Bc\frac{1}{1+K}.\end{equation} We formally obtain the vacuum solution for
$\lambda=1$.}, we can construct
the appropriate one parameter family of pure gauge solutions, $\Psi_\lambda$:
\begin{equation}\Psi_\lambda = \lambda \Psi -\lambda(1-\lambda)\left(
cB\frac{1+K}{1-\lambda+K}c \frac{1}{1+K}\right),\end{equation}
where $\Psi$ is the vacuum solution \EQ{solution}. Assuming the second term
vanishes as $\lambda$ approaches $1$, the vacuum and pure gauge solutions
appear to become identical. But we should be more careful.
Using the Schwinger representation to expand the second term more explicitly:
\begin{equation}\lim_{\lambda\to1}(\Psi-\Psi_\lambda)
= c B(1+K) \lim_{\lambda\to1}\left[(1-\lambda)\int_0^\infty dt\,
e^{-(1-\lambda)t}\Omega^t\right] c\frac{1}{1+K} .\label{eq:temp}\end{equation}
In this form the subtlety of the limit is clear. Though $1-\lambda$ vanishes,
as $\lambda\to 1$ there is a corresponding divergence from the integration
over all wedge states ($\Omega^t$ approaches a constant---the sliver
state---for large $t$). The product of these factors is finite, and in fact
\begin{equation}\lim_{\lambda\to 1^-} (1-\lambda)\int_0^\infty dt\,
e^{-(1-\lambda)t}\,\Omega^t =\Omega^\infty,\end{equation}
where $\Omega^\infty$ is the sliver state. Substituting into \EQ{temp}
therefore gives\footnote{We ignore the $1+K$ factor since this would give
a subleading contribution to the phantom piece, though such contributions can
be important \cite{Erler3}.}
\begin{equation}\lim_{\lambda\to1}(\Psi-\Psi_\lambda)
= c B\Omega^\infty c\frac{1}{1+K}.\end{equation}
Since $B$ annihilates the sliver when contracted with Fock space
states \cite{Schnabl,Erler2}, the last term is a phantom piece. However,
unlike in $\B$ gauge, the phantom term appears in the pure gauge solution (as
$\lambda$ approaches $1$), not the tachyon vacuum.

\subsection{Closed String Tadpole}
\label{subsec:Ian}

Since our solution describes an empty vacuum without D-branes, the field
configuration should leave the closed string background undisturbed. One way
to check this is to compute the closed string tadpole, which can be
evaluated as a disk amplitude
\begin{equation}\mathcal{A}_\Phi(\mathcal{V})=
-\langle \mathcal{V}(i\infty)
c(0)\rangle_{C_1,\mathrm{BCFT}_\Phi}.\end{equation}
Here $\mathcal{V}=c\tilde{c}\mathcal{V}^m$ is an on-shell closed string vertex
operator, and for convenience we have mapped the canonical unit disk to a
cylinder $C_1$ of unit circumference; the subscript $\mathrm{BCFT}_\Phi$
indicates that the correlator is evaluated in the boundary conformal field
theory corresponding to the classical solution $\Phi$.
Ellwood \cite{Ellwood} gave a nice prescription
for computing this amplitude directly from $\Phi$:
\begin{equation}\mathcal{A}_\Phi(\mathcal{V}) =
\mathcal{A}_0(\mathcal{V})+\lbra V\Phi\rbra,\end{equation}
where $\mathcal{A}_0(\mathcal{V})$ is the tadpole in the
reference BCFT defining the string field theory, and
$V=\mathcal{V}(i)|I\rangle$ \footnote{$\lbra V\Phi\rbra$
are the gauge invariant overlaps introduced in
\cite{Shapiro,Itzhaki,VSFT}.}.
This quantity is very easy to compute. The BRST exact term in \EQ{solution}
does not contribute, so we have
\begin{equation}\lbra V\Psi\rbra= \Llbra Vc\frac{1}{1+K}
\Rrbra =\int_0^\infty dt\, e^{-t}\lbra Vc\Omega^t\rbra.
\end{equation}
The inner product $\lbra Vc\Omega^t\rbra$ is a correlator on a cylinder of
circumference $t$; by a scale transformation we can reduce it to a cylinder of
unit circumference, producing a factor of $t$ for the $c$ ghost from the
conformal transformation. Thus
\begin{eqnarray}\lbra V\Psi\rbra =\lineup \lbra Vc\Omega\rbra
\int_0^\infty dt\,te^{-t}= \lbra Vc\Omega\rbra\nonumber\\
= \lineup \langle \mathcal{V}(i\infty)c(0)\rangle_{C_1} =
-\mathcal{A}_0(\mathcal{V}).\end{eqnarray}
Therefore the closed string tadpole vanishes:
\begin{equation}\mathcal{A}_\Psi(\mathcal{V})=0.\end{equation}
It is interesting to note that for our solution the contribution to the
amplitude comes from the BRST nontrivial term $c\frac{1}{1+K}$, whereas in
$\B$ gauge it comes exclusively from the phantom piece \cite{Ellwood}.

Before concluding, let us mention that it is possible to generalize this
calculation by computing the full off-shell boundary state of our solution,
following \cite{KOZ}. The calculation would take us too far astray to present
here, but we have confirmed that the boundary state for our solution vanishes
identically.

\section{Level Expansions}
\label{sec:level}

Though we have a simple analytic proof of Sen's first conjecture, it is
desirable to confirm our calculation by other means. The most trusted---but
also the most poorly understood---method for calculating the energy is the old
$L_0$ level expansion, which provided the first convincing numerical evidence
for Sen's conjectures in \cite{K&S,SZ,Moeller,Rastelli}. The level
expansion of our new solution, however, brings a surprise: if we add
contributions to the energy level by level, the expansion is divergent.

The situation here appears to be analogous to the ``sliver frame''
$\mathcal{L}_0$ level expansion, where the energy is represented as the formal
sum of an asymptotic series \cite{Schnabl,Aldo}. For our new solution, the
$\mathcal{L}_0$ level expansion is so simple that we are able to find an exact
expression for the asymptotic series and its resummation, allowing us to gain
concrete insight into the nonperturbative structure of the level expansion. The
$L_0$ case, of course, is more complicated, but we have found a useful toy
model of our solution where, remarkably, it is possible to compute the $L_0$
level expansion exactly in terms of elliptic functions. In both $L_0$ and
$\mathcal{L}_0$ expansions, we resum the divergent series to obtain good
agreement with Sen's first conjecture.

\subsection{Curly $\mathcal{L}_0$ Level Expansion}
\label{subsec:Curly}

We begin by considering the $\mathcal{L}_0$ level expansion. The
$\mathcal{L}_0$ level expansion is quite analogous to the ordinary $L_0$ level
expansion, but performed in a conformal frame well-adapted to the wedge
state geometry of analytic solutions. $\mathcal{L}_0$ is the dilatation
generator in the sliver conformal frame \cite{Schnabl}:
\begin{eqnarray}\mathcal{L}_0 = \lineup f_\mathcal{S}^{-1} \circ L_0
\nonumber\\ =\lineup \oint_0
\frac{d\xi}{2\pi i}(1+\xi^2)\tan^{-1}\xi\, T(\xi),\end{eqnarray}
where $f_\mathcal{S}(z)=\frac{2}{\pi}\tan^{-1}z$ is the sliver coordinate map.
Here, we define a state to be at level $L$ if it is an eigenstate of
$\mathcal{L}_0$ with eigenvalue $L$. We write such states in the form
\begin{equation}F\phi F,\end{equation}
where $F=\sqrt{\Omega}$ is the square root of the $SL(2,\mathbb{R})$ vacuum,
and $\phi$ corresponds to an insertion of an operator with scaling dimension
$L$ in the sliver coordinate frame. $K,B,c$ have scaling dimension $1,1,-1$
respectively, and the dimensions are additive with the star product.
Therefore, any state at level $L$ in the $KBc$ subalgebra can be written using
states of the form
\begin{equation}F \Big(K^l cB K^m cK^n\Big) F,\ \ \ \ \ l+m+n = L+1.
\label{eq:L0_eig}\end{equation}
This is a different basis of eigenstates from the one used in
\cite{Schnabl}, but either basis gives the same level expansion for the
energy.

To expand the solution (\ref{eq:real_sol}) in terms of $\mathcal{L}_0$ 
eigenstates, we multiply and divide by $F$,
\begin{equation}\hat{\Psi}=
F\left(\frac{e^{K/2}}{\sqrt{1+K}}\Big[c+cKBc\Big]
\frac{e^{K/2}}{\sqrt{1+K}}\right)F,\end{equation}
and expand the factor in parentheses in powers of $K$. It is useful to
introduce the field
\begin{equation}\hat{\Psi}(z) = z^{\mathcal{L}_0}\hat{\Psi} =
F\left(\frac{e^{zK/2}}{\sqrt{1+zK}}\Big[\frac{1}{z}c+cKBc\Big]
\frac{e^{zK/2}}{\sqrt{1+zK}}\right)F.\label{eq:psiz}\end{equation}
Then the $\mathcal{L}_0$ level expansion is equivalent to a power series
expansion in $z$. Note that in our convention the expansion starts at level
$-1$ with the zero-momentum tachyon $FcF=\frac{2}{\pi}c_1|0\rangle$.

To compute the energy we should sum the infinite series
\begin{equation}E = \sum_{n=-2}^\infty E_n ,\label{eq:E_sum}\end{equation}
where $E_n$ is the contribution to the energy (or the action) coming from
fields whose levels add up to $n$. Assuming the equations of motion, the $E_n$s
can be found from the expression
\begin{equation}E_n = \frac{1}{6}\oint_0 \frac{dz}{2\pi i}\frac{1}{z^{n+1}}
\langle\hat{\Psi}(z),Q_B\hat{\Psi}(z)\rangle.\end{equation}
Therefore, to find the expansion we should evaluate the inner product
\begin{equation}
E(z)=\frac{1}{6}\langle\hat{\Psi}(z),Q_B\hat{\Psi}(z)\rangle.\end{equation}
In $\B$ gauge, the computation of this quantity appears to be a nontrivial
task, but for our new solution it is
quite straightforward. The final answer is naturally expressed in terms of a
variable $Z$, related to $z$ by an $SL(2,\mathbb{R})$ transformation:
\begin{equation}Z = \frac{1}{2}\frac{z}{1-z}.\end{equation}
We find
\begin{equation}E(z)=-\frac{1}{2\pi^2}\left[1+\frac{2}{3}\frac{1}{Z}
+\frac{1}{6}\frac{1}{Z^2}+\frac{1}{6\pi}\frac{I(Z)}{Z^4}\right],
\label{eq:Ez}\end{equation}
where $I(Z)$ is the integral\footnote{$I(Z)$ can actually be expressed
in terms of a known function, called the incomplete Bessel function
\cite{Bessel}}
\begin{equation}
I(Z)=\int_0^\infty du\, e^{-u/Z}
(u+1)^3\sin\frac{\pi}{u+1}.\end{equation}
Note that as $z$ approaches $1$ (or $Z\to\infty$) the energy
function approaches the expected value $E(1)=-\frac{1}{2\pi^2}$.

To find the $E_n$s, we need a power series expansion for this integral. To this
end, expand the second factor in the integrand as a Taylor series:
\begin{equation}(1+u)^3\sin\frac{\pi}{1+u}= \sum_{n=1}^\infty \ell_n u^n,
\end{equation}
where $\ell_n$s can be expressed in terms of generalized Laguerre polynomials
\begin{equation}\ell_n =
(-1)^n\mathrm{Im}\big[L_n^{-4}(i\pi)\big].\end{equation}
Integrating over $u$ produces a factor of $n!$ in the sum, so we find the
power series for $E(z)$
\begin{equation}E(z)=-\frac{1}{2\pi^2}\left[1+\frac{2}{3}\frac{1}{Z}
+\frac{1}{6}\frac{1}{Z^2}+\frac{1}{6\pi}\sum_{n=1}^\infty n!\ell_n\,Z^{n-3}
\right].\label{eq:cLexp}\end{equation} This is a prototype for an asymptotic
expansion. The $n!$ divergence of the coefficients is not helped by the
$\ell_n$s,
which themselves diverge quite rapidly\footnote{The large $n$ asymptotics of
the Laguerre polynomials implies $\ln|\ell_n|\sim \sqrt{2\pi n}$. We have
confirmed this behavior numerically.} due to the essential singularity in the
Laguerre generating function at $u=-1$.

From here it is a trivial extra step to expand $Z$ in terms of
$z$ and read off the $E_n$s. To the first few orders, we find explicitly:
\begin{eqnarray}E(z) =\lineup \frac{1}{6}
\left[-\frac{4}{\pi^2}\frac{1}{z^2}+\left(-\frac{2}{\pi^2}+\frac{1}{2}\right)
-\frac{\pi^2}{8}z^2+\frac{\pi^2}{2}z^3+\left(-\frac{33\pi ^2}{16}
+\frac{\pi^4}{32}\right) z^4+\left(\frac{37 \pi ^2}{4}-\frac{3\pi^4}{8}\right)
z^5\right.\nonumber\\ \lineup\ \ \ \ \ \ \ \
\left.+\left(-\frac{365 \pi ^2}{8}+\frac{55 \pi^4}{16}-\frac{\pi^6}{128}
\right) z^6+\left(\frac{987 \pi ^2}{4}-\frac{235 \pi ^4}{8}+\frac{3\pi ^6}{16}
\right) z^7+...\right].\end{eqnarray}
This gives an efficient method for computing $E_n$s. Indeed, we were easily
able to compute the $E_n$s out to $n=400$ and could have gone much further,
whereas with our current understanding the calculation in $\B$ gauge becomes
time consuming much beyond $n=50$.

\begin{table}[t]
\begin{center}
\begin{tabular}{|c|c|c|c|c|c|c|}\hline
$N$ & -2 & 0 & 2 & 4 & 6 & 8  \\ \hline
New solution & $-1.3333$ & $-0.35507$ & $-4.4137$ & $-45.133$ & $-269.51$ &
$22051$ \\ \hline
$\B$ gauge & $-1.3333$ & $-1.0015$ & $-0.98539$ & $-1.0327$ & $-1.3054$ &
$6.7582$ \\ \hline
\end{tabular}
\end{center}
\caption{\label{tab:part_sum} Partial sum $\sum_{n=-2}^N E_n$ up
to $N=8$ in units of $\frac{1}{2\pi^2}$, shown for the new solution
\EQ{solution},\EQ{real_sol} and the $\B$ gauge solution, taken
from \cite{Schnabl}.}
\end{table}

For illustrative purposes, we have listed the first few partial sums of the
$E_n$s in table \ref{tab:part_sum}, both for the new solution and the $\B$
gauge solution. Both reveal an ``approximation'' to the energy which is typical
of a divergent asymptotic series. However, the partial sum for our new solution
diverges much faster than in $\B$ gauge---ironically, the best approximation to
the energy is the trivial one, where we truncate the solution down to the zero
momentum tachyon.

To compute the energy, it is necessary to resum the asymptotic series.
One way to do this is to use the method of
Pad\'e approximants \cite{Schnabl}, where we replace the asymptotic series
$z^2E(z)$ by a Pad\'e approximant $P_m^n(z)$---a ratio of a degree $n$
polynomial to a degree $m$ polynomial chosen so that the first $m+n$ terms in
the Taylor expansion of $P_m^n(z)$ match those of $z^2E(z)$. The
approximation to the energy is then revealed by evaluating $P_m^n(1)$. A second
method\footnote{We thank D. Gross for suggesting this to us.} is to use a
combination of Pad\'e and Borel resummation. Here we replace the Borel
transform of $z^2E(z)$ by its Pad\'e approximant $P_m^n(z)_{\mathrm{Borel}}$
and evaluate the integral
\begin{equation}\widetilde{P}_m^n(z)=
\int_0^\infty dt\, e^{-t} P_m^n(tz)_{\mathrm{Borel}}
\end{equation}
at $z=1$. In table \ref{tab:resum} we list Pad\'e and Pad\'e-Borel
approximations to the energy including fields out to level $199$. Both
confirm Sen's conjecture to very high accuracy. At low levels, Pad\'e-Borel
does a little better than Pad\'e, though at very high levels Pad\'e appears
to be more accurate\footnote{Note that the
convergence is slower than it is in $\B$ gauge: to get results as good as our
$P_{60}^{60}(1)$, one only has to go out to $P_{18}^{18}(1)$ in $\B$ gauge.}.

\begin{table}[t]
\renewcommand{\arraystretch}{1.0}
\begin{center}
\begin{tabular}{lclcl}
& $\ \ \ \ \ \ \ \ \ \ \ \ \ $
& $P_n^n(1)$
& $\ \ \ \ \ \ \ \ \ \ \ \ \ $
& $\widetilde{P}_n^n(1)$ \\ \hline
$n=0$ & & $-1.33333$ & & $-1.33333$\\
$n=2$ & & $-1.14334$ & & $-0.994896$\\
$n=4$ & & $-0.898883$ & & $-0.900412$\\
$n=6$ & & $-1.04241$ & & $-1.00487$\\
$n=8$ & & $-0.996478$ & & $ -1.00029$\\
$n=10$ & & $-0.995773$ & & $-0.999944$\\
%$n=12$ & & $-0.999000$ & & $-1.00005$\\
%$n=14$ & & $ -0.999417$ & & $-1.00009$\\
%$n=16$ & & $ -0.999199$ & & $-0.999821$\\
%$n=18$ & & $ -0.999907$ & & $-0.999962$ \\
$n=20$ & & $ -0.99991237$ & & $-0.99996793$ \\
$n=40$ & & $ -0.99998202$ & & $-0.99999517$ \\
$n=60$ & & $ -0.99999945$ & & $-0.99999754$ \\
$n=80$ & & $ -0.99999984$ & & $-0.99999904$ \\
$n=100$ & & $ -0.99999995$ & & $-0.99999954$ \\ \hline
\end{tabular}
\end{center}
\caption{\label{tab:resum} Pad\'e and Pad\'e-Borel approximation to the energy
in units of $\frac{1}{2\pi^2}$. We have
shown the approximants for $m=n$, since Pad\'e resummation is generally most
reliable when the numerator and denominator are polynomials of similar order.
%We are confident of these numbers out to 300 digits of precision, though we
%have not displayed them all here.
}
\end{table}

\begin{figure}
\begin{center}
\resizebox{3.1in}{2.5in}{\includegraphics{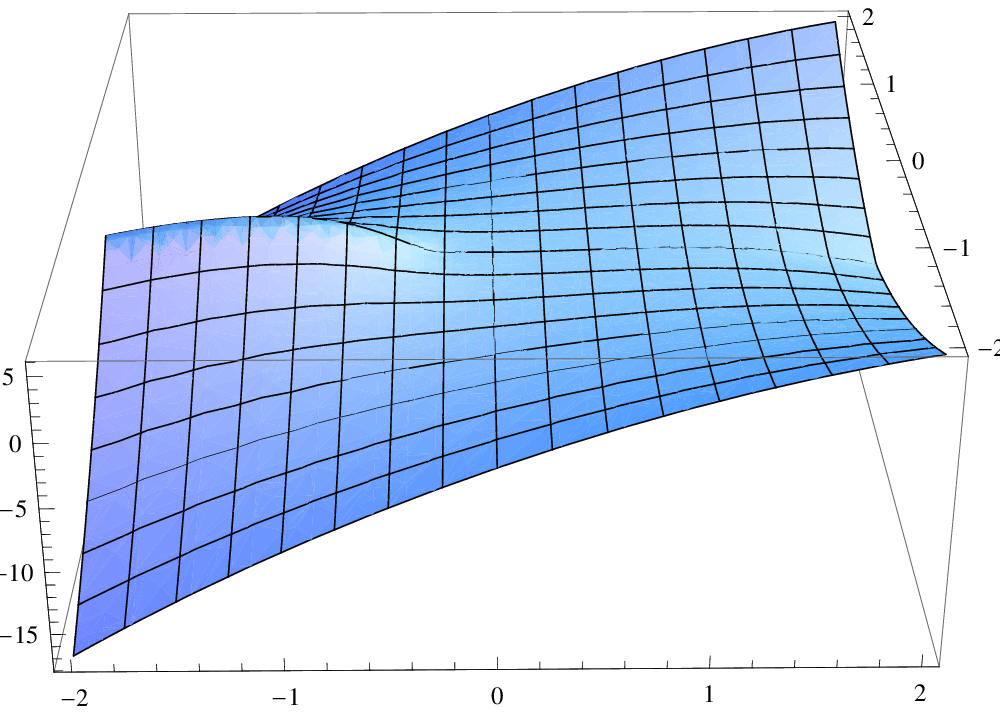}}
\resizebox{3.1in}{2.5in}{\includegraphics{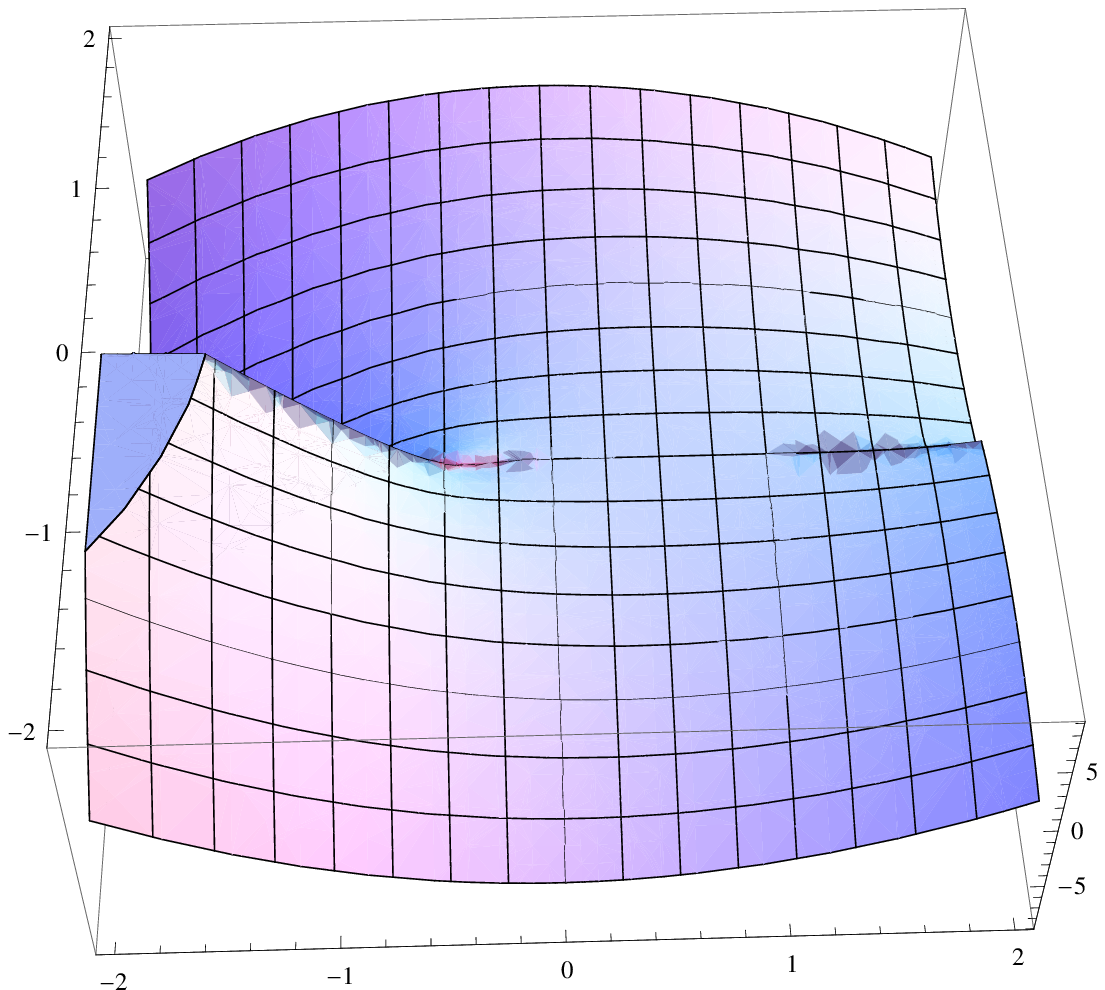}}
\end{center}
\caption{\label{fig:graphs} Real and imaginary parts of
$z^2E(z)$ for $-2<\mathrm{Re}(z)<2$ and $-2<\mathrm{Im}(z)<2$, shown left and
right, respectively. Note that the function is very smooth at $z=0$ and $1$,
but they are nevertheless branch points.}\end{figure}

It is interesting to understand why the $\mathcal{L}_0$ level expansion is
asymptotic. By analogy with the old argument about the divergence of
perturbation theory in QED, one suspects that something severe must happen
to the energy $E(z)$ as the ``coupling constant'' $z$ is taken to be negative.
The problem is easy to identify: for $z<0$ the string field $\hat{\Psi}(z)$
does not exist. That is, though $\hat{\Psi}(z)$ has a
well-defined expansion in terms of $\mathcal{L}_0$ eigenstates, for $z<0$ the
expansion does not converge to a well defined string field. The
problem comes from the factor $\frac{1}{1+zK}$, which for
$z<0$ would only seem to make sense as an integral over singular ``inverse''
wedge states. This fact should show up as some sort of pathology in the
energy $z^2E(z)$ for $z\leq0$. In fact, because we have a closed form
expression \EQ{Ez}, we can plot the energy to see what happens. As can be
seen from figure \ref{fig:graphs}, $z^2E(z)$ has a branch point at $z=0$
together with a branch cut extending to $z=\infty$.
Though we can analytically continue to negative $z$, the continuation is
not unique and moreover is {\it complex}, in contradiction with the fact
that $\hat{\Psi}(z)$ is real to any finite level in the level expansion.
Therefore $z^2 E(z)$ for $z<0$ cannot be interpreted as a BRST inner product
of $\hat{\Psi}(z)$. Incidentally, note that there is another branch point at
$z=1$. This comes from the factor $Fe^{zK/2}$, which for $z>1$ is an inverse
wedge state.

We expect that this phenomenon is quite general. For any solution
depending on some $f(K)$ expressed in terms of positive powers of the
$SL(2,\mathbb{R})$ vacuum, $f(zK)$ for $z<0$ will be undefined. Therefore
the energy function should be singular at $z=0$, rendering the
$\mathcal{L}_0$ level expansion asymptotic.

\subsection{Square $L_0$ Level Expansion}
\label{subsec:Square}

The traditional $L_0$ expansion of a string field very efficiently summarizes
all possible overlaps with Fock states up to a given conformal weight. Such an
information is often useful, either in explicit numerical computations, or as
one possible criterion of a string field being well defined.

To expand our solution in the eigenstates of $L_0$ it is convenient to use
the techniques and formalism of \cite{Schnabl}.  The twist even (real) solution
can be written as
\be\label{eq:Psi_L0ready}
\hat\Psi=\frac{1}{\pi} \int_0^\infty \!\!\! \int_0^\infty dt \, ds\,
\frac{e^{-t-s}}{\sqrt{t s}} \Uhat_{t+s+1} \left[\frac{2}{\pi} \tilde c
\left(\frac{\pi}{4} (s-t) \right)  +\frac{1}{\pi} Q_B \Bhat \tilde c
\left(\frac{\pi}{4} (s-t) \right)\right] \ket{0},
\ee
where $\Uhat_r = U_r U_r^\star$ and the star denotes the BPZ conjugate. The
rest of the notation follows \cite{Schnabl}, in particular $U_r =
(2/r)^{\ll_0}$. The tilde is used to translate the $c$ insertions in the
cylinder frame to the canonical upper half plane, explicitly $\tilde c(x) =
\cos^2 x \, c(\tan x) $.

The string field can be readily expanded and the individual coefficients can be
numerically integrated. We find
\bea\label{eq:Psi_L0}
\hat\Psi &=& 0.509038 \, c_1 \ket{0}+ 0.13231 \, c_{-1} \ket{0}  -0.00157618 \, L_{-2} \, c_1 \ket{0}  +\nonumber\\
&& -0.0135795 \, L_{-4}\, c_1 \ket{0} +0.0231579 \, L_{-2} L_{-2}\, c_1 \ket{0} + 0.0893356 \, c_{-3} \ket{0}  \nonumber\\
&& -0.00694698 \, L_{-2} \, c_{-1} \ket{0} + \cdots + (Q_B \mbox{-exact}).
\eea
For example the first coefficient is given by
\begin{eqnarray}
t &=& \frac{1}{2\pi^2} \int_0^\infty du \int_{-1}^1 dw \, e^{-u}
\frac{(u+1)^2}{\sqrt{1-w^2}} \cos^2 \left(\frac{\pi}{2} \frac{u}{u+1}w\right)
\nonumber\\
&=& \frac{1}{4\pi} \int_1^\infty du \, e^{1-u} u^2 \left(1+J_0\left(\pi
\frac{u-1}{u}\right)\right)
\nonumber\\
&=& 0.509038,\label{eq:t_coeff1}
\end{eqnarray}
where $J_0$ is a Bessel function of the first kind. To obtain \EQ{t_coeff1}
from \EQ{Psi_L0ready} we have made a change of variables $u=t+s$ and
$w=(t-s)/(t+s)$. In more generality all the coefficients are given by an
integral of the form
\begin{equation}
\int_0^\infty du  (u+1)^2 P\left(\frac{1}{u+1}\right) e^{-u} \int_{-1}^1 dw
\frac{1}{\sqrt{1-w^2}} \cos^2 \left(\frac{\pi}{2} \frac{u}{u+1}w\right) \tan^n
\left(\frac{\pi}{2} \frac{u}{u+1}w\right),
\end{equation}
where $P$ is a polynomial whose detailed form depends on the coefficient in
question. These integrals are absolutely convergent, but to
evaluate them numerically with enough precision we found necessary to make a
further change of variables $w=\sin\phi$ upon which the integrable singularity
at $w=\pm 1$ disappears.

The apparently rapid decay of the coefficients suggests that the energy of the
solution computed in level truncation should converge quite well. Let us
compute the regularized energy, the analogue of \EQ{Ez}:
\be
\label{eq:EL_0}
\widetilde{E}(z) = \frac{1}{6} \langle z^{L_0}\hat\Psi, Q_B z^{L_0}
\hat\Psi\rangle.
\ee
For $z=1$ we recover the exact expression, and because the kinetic term is
diagonal in $L_0$ eigenstates, the coefficients of the energy at order
$z^{2L-2}$ are exactly the contributions from fields at level $L$. (Here, 
following usual convention, the level refers to the eigenvalue of $L_0+1$.)
With the help of the computer\footnote{Part of our computer code was written
by Ian Ellwood while working on an unpublished project with the second author
\cite{multiple}. We thank him for kindly letting us use his code.} we have
computed the energy up to level 30 which in our basis includes contributions
from 2455 fields.  The resulting (normalized) energy takes the form
\begin{eqnarray}\label{eq:sqEz}
2\pi^2 \widetilde{E}(z) =\lineup -\frac{0.85247}{z^2} - 0.0616762 z^2 - 0.120529 z^6 +
0.104037 z^{10} -
 0.132712 z^{14} + 0.158365 z^{18}
\nonumber\\
 \lineup- 0.204746 z^{22} + 0.268088 z^{26} -
 0.363999 z^{30} + 0.496009 z^{34} - 0.682054 z^{38} +
\nonumber\\
 \lineup +0.942044 z^{42} - 1.30865 z^{46} + 1.81739 z^{50} - 2.52216 z^{54}
+ 3.49649 z^{58}+ \cdots .
\end{eqnarray}
The result for the lowest levels is encouraging: at lowest truncation level we
find 85\% of the expected energy, at level 2 we get 91\% and at level 4 already
103\%.  But that is as close as we get to the right answer; in fact it is
obvious from \EQ{sqEz} that the contributions of higher levels are
increasing in magnitude and therefore the series cannot converge.

As we've seen, a similar divergence occurs in the $\mathcal{L}_0$ level
expansion, but this is the first time such behavior has appeared in the
canonical $L_0$ level truncation scheme. We can evaluate the energy using
either Pad\'{e} or Pad\'{e}-Borel resummation; as shown in table
\ref{tab:resum_2}, both types of resummation confirm Sen's conjecture to better
than 99\% at level 30. It is of great interest to understand why the 
expansion of our solution is divergent. We explore the answer to this question
using an explicitly soluble toy model in section \ref{subsec:Id_corr}.

\begin{table}
\renewcommand{\arraystretch}{1.0}
\begin{center}
\begin{tabular}{lclcl}
& $\ \ \ \ \ \ \ \ \ \ \ \ \ $ & $P_n^n(1)$ & $\ \ \ \ \ \ \ \ \ \ \ \ \ $ &
$\widetilde{P}_n^n(1)$ \\ \hline
$n=0$ & & $-0.852470$ & & $-0.852470$\\
$n=4$ & & $-0.787834$ & & $-0.871988$\\
$n=8$ & & $-0.992052$ & & $-0.983243$\\
$n=12$ & & $-0.992013$ & & $-0.984516$\\
$n=16$ & & $-0.996081$ & & $ -0.993936$\\
$n=20$ & & $-0.999595$ & & $-0.993687$\\
$n=24$ & & $-0.997322$ & & $-0.995001$\\
$n=28$ & & $-0.997690$ & & $-0.993253$\\
\hline
\end{tabular}
\end{center}
\caption{\label{tab:resum_2} Pad\'e and Pad\'e-Borel approximation to the
energy in units of $\frac{1}{2\pi^2}$. We have shown the approximants for
$m=n$. Note that the approximants $P_n^n$ include the contributions of fields
up to level $n$. }
\end{table}

Let us give the expansion of our solution in the original matter Virasoro+ghost
oscillator basis used by Sen and Zwiebach \cite{SZ}, out to level 4:
\bea
\hat\Psi &=& t c_1 \ket{0} + u c_{-1} \ket{0} + v  L_{-2}^{m}\, c_1 \ket{0} + w  b_{-2} c_{0} c_{1} \ket{0} +\nonumber\\
&& + A  L_{-4}^{m}\, c_1 \ket{0} + B L_{-2}^{m} L_{-2}^{m}\, c_1 \ket{0} + C c_{-3} \ket{0} + D  b_{-3} c_{-1} c_{1} \ket{0}+ \nonumber\\
&& + E  b_{-2} c_{-2} c_{1} \ket{0} + F  L_{-2}^{m} c_{-1} \ket{0} +  w_1 L_{-3}^{m} c_0 \ket{0} + w_2 b_{-2} c_{-1} c_{0} \ket{0} +\nonumber\\
&&  + w_3 b_{-4} c_{0} c_{1} \ket{0} + w_4  L_{-2}^{m} b_{-2} c_{0} c_{1}
\ket{0} + \cdots.
\eea
The coefficients above are given by
\bdm\small
\begin{array}{|cclcclcclccl|}
\hline
t &=&   0.509038 & \quad A &=&  -0.10674   & \quad E &=&   0.242131  & \quad
w_1 &=& 0
\nonumber \\
u &=&   0.772988   & \quad B &=&   0.106714  & \quad F &=&   0.673728 & \quad
w_2 &=& 1.13718
\nonumber \\
v &=&   0.213559   & \quad C &=&  1.11009    &         & &               &
\quad w_3 &=& 0.3338
\nonumber \\
w &=&  -0.211983  & \quad D &=&   0.887287    &         & &               &
\quad w_4 &=& -0.343299.
\nonumber \\
\hline
\end{array}
\edm
Surprisingly, the expectation values do not appear to be getting smaller at
higher levels, at least out to level 4. Apparently this is an artifact of
the choice of basis, since in the simpler basis \EQ{Psi_L0} the coefficients
appear to decay quite rapidly. Of course, the level
approximation to the energy is the same in either case.

\begin{table}
\renewcommand{\arraystretch}{1.0}
\begin{center}
\begin{tabular}{lclcl}
& $\ \ \ \ \ \ \ \ \ \ \ \ \ $ & $P_n^n(1)$ & $\ \ \ \ \ \ \ \ \ \ \ \ \ $ &
$\widetilde{P}_n^n(1)$ \\ \hline
$n=0$ & & $-0.266085$ & & $-0.266085$\\
$n=4$ & & $-0.679355$ & & $-0.679026$\\
$n=8$ & & $-0.935655$ & & $-0.883524$\\
$n=12$ & & $-0.940574$ & & $-0.920585$\\
$n=16$ & & $-0.971911$ & & $ -0.950665$\\
$n=20$ & & $+0.452292$ & & $-0.946722$\\
$n=24$ & & $-0.974222$ & & $-0.955226$\\
$n=28$ & & $-0.974103$ & & $-0.954514$\\
\hline
\end{tabular}
\end{center}
\caption{\label{tab:resum_3} Pad\'e and Pad\'e-Borel approximation to the
energy for the asymmetric solution in units of $\frac{1}{2\pi^2}$. We have
shown the approximants for $m=n$.
% To obtain the $n=28$ approximants we had to evaluate the contributions to the energy up to level 28.
The value $P_{20}^{20}$ is anomalously large due to an accidental position of a
zero and a pole of the Pad\'e approximant very near the value $z=1$. }
\end{table}

It is of interest to consider the level expansion of the non-real solution
\EQ{solution} as well. Focusing on the BRST nontrivial part of the string
field we find by numerical integration
\bea\label{eq:Psia_L0}
\Psi &=& 0.284394 \, c_1 \ket{0}+ 0.249034 \, c_0 \ket{0} + 0.244516 \, c_{-1} \ket{0}  +0.0359031 \, L_{-2} \, c_1 \ket{0}  +\nonumber\\
&& +0.252567 \, c_{-2} \ket{0} + 0.00302175 \, L_{-2}\, c_0 \ket{0} -0.0177251
\, L_{-4}\, c_1 \ket{0} + \\ \nonumber
 && +0.0175741 \, L_{-2} L_{-2}\, c_1 \ket{0} + 0.268936\, c_{-3} \ket{0} -0.010923 \, L_{-2} \, c_{-1} \ket{0} + \cdots +
(Q_B \mbox{-exact}).
\eea
We have computed the components of the string field up to level 30. The
resulting $z$-dependent energy is given by
\begin{eqnarray}\label{eq:sqEza}
2\pi^2 \widetilde{E}_{\mathrm{asym}}(z)
&=& -\frac{0.266085}{z^2}-0.408062 -0.00644403 z^2
+0.0200865 z^4-0.292541z^6-0.108361 z^{8} \nonumber\\
&& + 0.23035 z^{10}+0.0672657 z^{12}-0.275233 z^{14}-0.074523 z^{16}+0.299372
z^{18} \nonumber\\
&&+0.0574889 z^{20} -0.362862 z^{22}-0.0592361 z^{24}+0.440743 z^{26}+0.0513536
z^{28}\nonumber\\
&&-0.563397 z^{30}-0.0524896 z^{32}+0.721687 z^{34}+0.0471252 z^{36}-0.944548
z^{38}\nonumber\\
&&-0.0474732 z^{40}+1.24749 z^{42}+0.0439229 z^{44}-1.67218 z^{46}-0.0442855
z^{48}\nonumber\\
&& +2.25055 z^{50}+0.0415004 z^{52}-3.04491 z^{54}-0.0416184 z^{56}+4.13094
z^{58}.
\end{eqnarray}
There are twice as many terms here because the solution is not twist
even, so odd levels contribute to the action as well. Again the expansion is
divergent and we can resum the series using Pad\'e or Pad\'e-Borel resummation.
The results in table \ref{tab:resum_3} nicely confirm
Sen's conjecture, though we do not get quite as close to the expected answer
as with the real solution.

\subsection{Exactly Soluble Model for the $L_0$ Level Expansion}
%\subsection{Identity Correlators in Level Truncation}
\label{subsec:Id_corr}

\begin{figure}
\begin{center}
\resizebox{2in}{2in}{\includegraphics{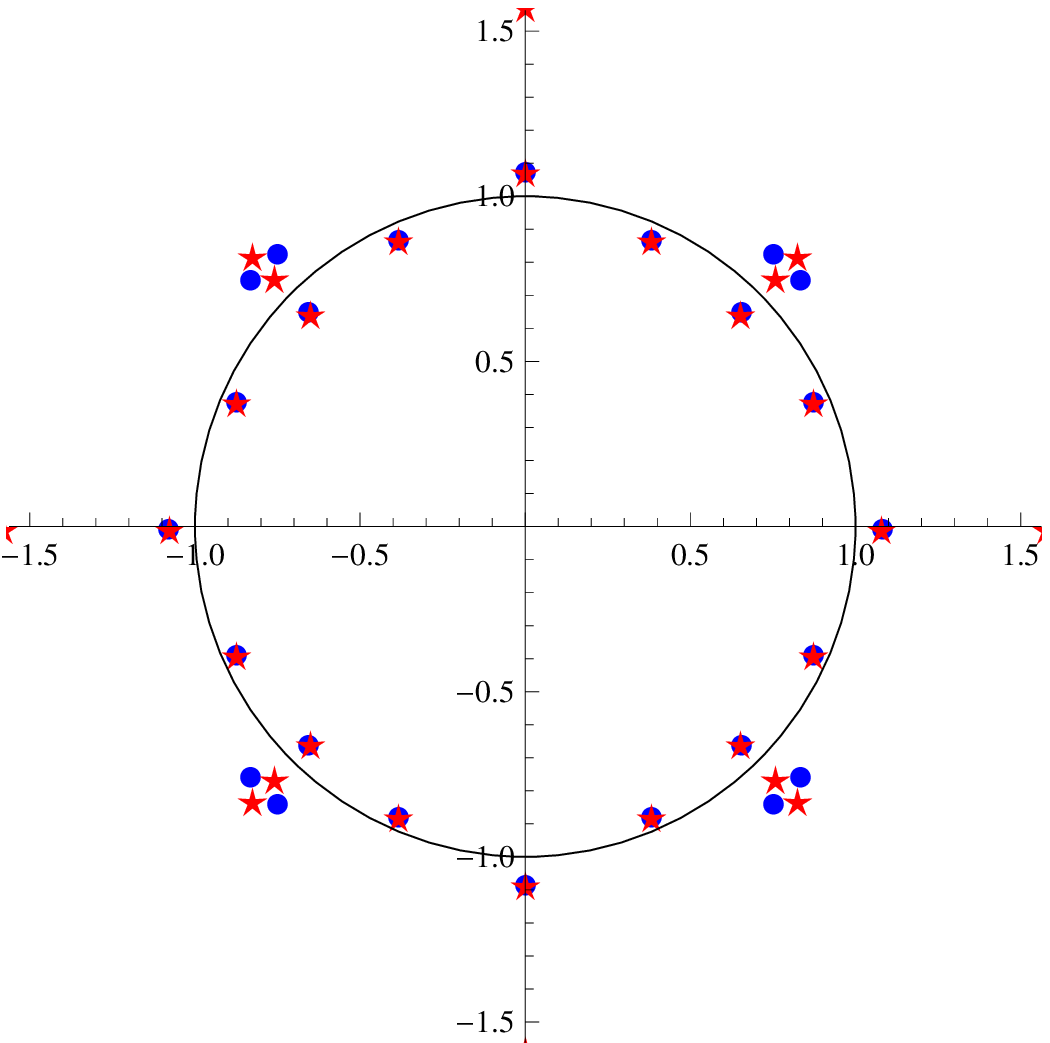}} $\qquad\qquad$
\resizebox{2in}{2in}{\includegraphics{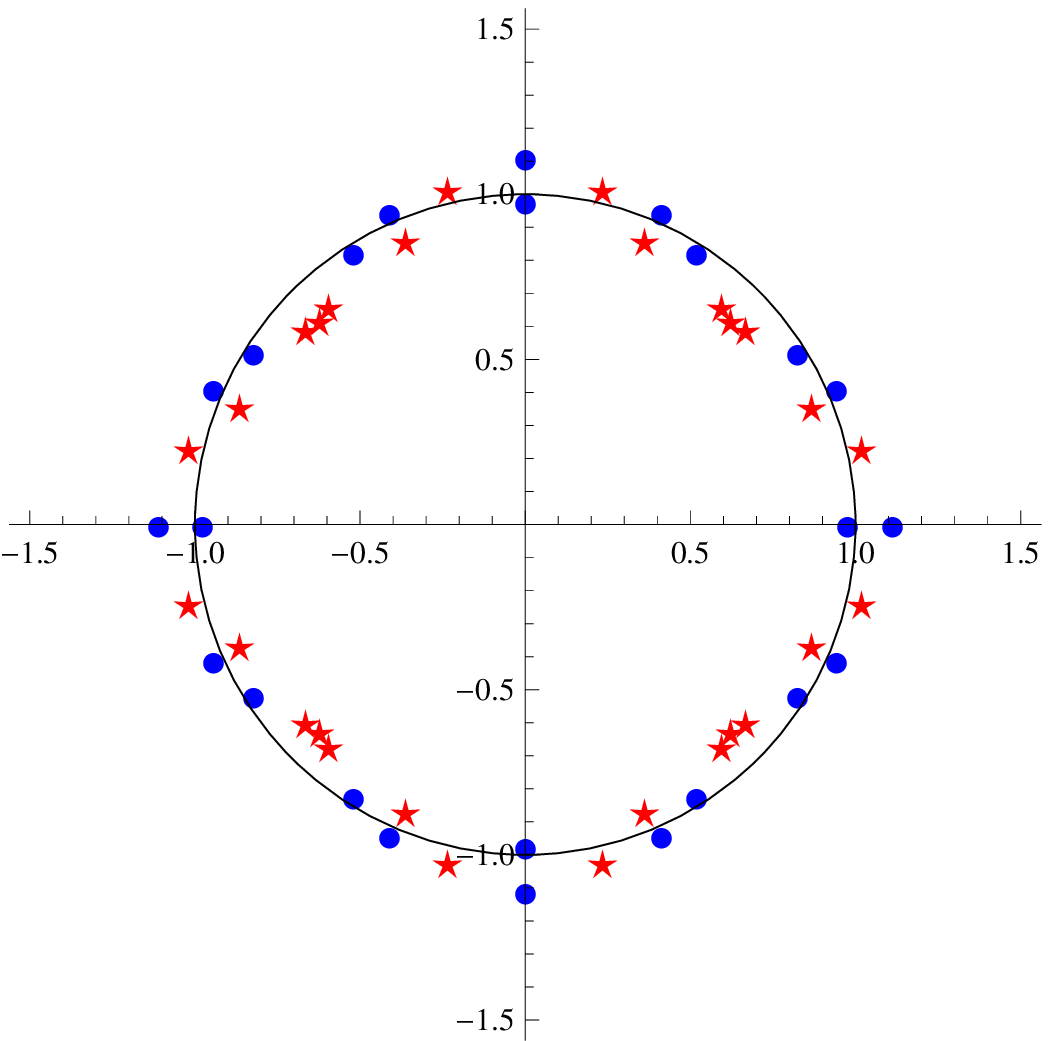}}
\end{center}
\caption{\label{fig:Pade_poles} a) Location of the poles and zeros of the
Pad\'e approximant $P^{30}_{30}$ of $z^2\widetilde{E}(z)$ in \EQ{EL_0}. Red
asterisks indicate position of poles; blue dots indicate location of zeros. b)
The analogous picture for the identity correlator
(\ref{eq:Polya}). Note that for the true solution the poles and zeros almost 
coincide, which suggests milder singularities along the unit circle than is 
present for the identity correlator.}\end{figure}

Let us now try to understand why the $L_0$ expansion of our solution is
divergent. Following the logic of section \ref{subsec:Curly}, the divergence
should be related to the analytic structure of the energy as a function of
the parameter $z$. Given the slow non-exponential growth of the coefficients
in \EQ{sqEz} we expect the function $z^2\widetilde{E}(z)$ to be holomorphic inside the
unit disk but with some singularities on its boundary. Plotting the
distribution of poles and zeros of Pad\'{e} approximants suggests that
$z^2\widetilde{E}(z)$ cannot be analytically continued beyond the unit disk, just like
elliptic functions in the $q$ variable (see figure \ref{fig:Pade_poles}).

% They decay much less rapidly than they do in the
%$\bb_0$-gauge solution \cite{Schnabl} which decay as
%$(-1/3)^n/n! e^{-1.21 n^{1/3}}$, resembling more the behavior of
%the sliver state.

We can gain an important insight into this problem by looking at a certain
class of coefficients in \EQ{Psi_L0}. For example the family of states
$(L_{-2})^n c_1 \ket{0}$ comes with coefficients given by
\be
v_n = \frac{(-3)^{-n}}{\pi (n - 1)!} \int_0^\infty \! du \,  e^{-u}
\left(1+J_0\left(\pi \frac{u}{u+1}\right)\right) \left(\frac{(u+3)(u-1)}{4n} -
\frac{2}{u+1}\right)\left(1-\frac{4}{(u+1)^2}\right)^{n-1}.
\ee
For large $n$, these behave as
\be
v_n =\frac{1}{2\pi n!}\left(1+O\left(\frac{1}{n}\right)\right),
\ee
This looks exactly as though the coefficients are coming from the
identity string field. This identity-like behavior is not surprising. The
dominant contribution to our solution comes from wedge states close to the
identity, since larger wedges are exponentially suppressed.

This suggests that we consider the field
$c =\frac{1}{\pi} U_1^*c_1|0\rangle$ as a simple toy model for the level
expansion of our solution $\hat{\Psi}$. The level expansion of $c$ will not
yield the brane tension ($c$ is not a solution), but it is of interest in its
own right in relation to certain other energy computations, as we will
describe shortly. The analogue of the $z$-dependent energy for $c$ is:
\be\label{eq:idcorr}F(z)=\left\langle z^{L_0}c,z^{L_0}
Q_B c\right\rangle
=\frac{1}{\pi^2}\aver{0|c_{-1} U_1 z^{2L_0} U_1^* c_1 c_0|0}.
\ee
To our great surprise, we found that the contribution to $F(z)$ from each
level is exactly an integer:
\bea\label{eq:Polya}
F(z)= -\lineup\frac{1}{4\pi^2}\left[ \frac{1}{z^2}-4 z^2 +10 z^6 -24z^{10} +55
z^{14} -116 z^{18} +230 z^{22} -\right. \nonumber\\
\lineup\ \ \ \ \ \
\left.-440 z^{26} +819 z^{30} -1480 z^{34}+2602 z^{38} + \cdots\right].
\eea
Such a nice expansion is sure to have an analytic explanation, but before
we derive it, let us note that the question about the analytic behavior of
$F(z)$ is essentially answered at this point. By the Polya-Carlson
theorem a function with integer coefficients in its Taylor
expansion cannot be extended beyond the unit disk unless it is rational
(which, as we will show, it is not). Therefore $F(z)$ must have an
essential singularity at every point on the unit circle. This agrees well with
the analytic structure $z^2\widetilde{E}(z)$ in \EQ{EL_0}, as suggested by
position of the Pad\'e poles and zeros.

\begin{figure}
\begin{center}
\resizebox{2.7in}{1.5in}{\includegraphics{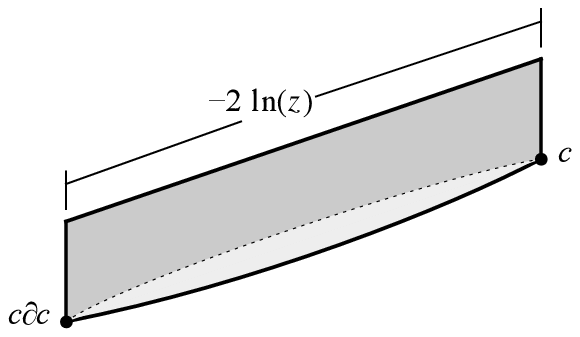}}
\end{center}
\caption{\label{fig:simple_strip} Worldsheet picture of our toy correlator
\EQ{idcorr}.}\end{figure}

Let us now see how to evaluate $F(z)$ analytically. Geometrically, \EQ{idcorr}
can be represented as a correlator of ghost operators on a paper-bag-shaped
surface obtained by taking a rectangular strip, folding it in half and gluing
together adjacent edges of the folded boundary (see figure
\ref{fig:simple_strip}). To evaluate the correlator directly one would have to
conformally map the geometry to the upper half plane where we know all the
correlation functions. Undoubtedly such a map can be constructed (along the
lines of \cite{Giddings})\footnote{Upon completion of this paper we were
informed by Ian Ellwood that such a map has been constructed in 
\cite{TZ,Garousi}.}, but there is a simple shortcut.

Algebraically, our task is to ``normal order'' $U_1 z^{2L_0} U_1^*$, that is,
to find a conformal map $\psi(\xi)$, holomorphic in the vicinity of
$\xi=0$ such that
\be\label{eq:1psieq}
U_1 z^{2L_0} U_1^* = U_\psi^* U_\psi,
\ee
where $U_\psi$ is the action of a finite conformal transformation $\psi(\xi)$ (note
that $\psi$ implicitly depends on $z$). If we can find such a $\psi$, then we can
easily compute $F(z)$:
\begin{equation}F(z) = -\frac{1}{\pi^2}\psi'(0)^{-2}.\end{equation}
In terms of conformal transformations the problem can be stated equivalently as
finding $\psi(\xi)$ holomorphic around the origin, such that
\be\label{eq:fpsieq}
f \circ I \circ f^{-1} \circ I = I \circ \psi^{-1} \circ I \circ \psi,
\ee
where $I$ stands for the inversion $I: \xi \to -1/\xi$, and $f$ is the map
entering the definition of the star algebra identity composed with rescaling by
$z$, $f(\xi)=\frac{2\xi}{1- z\xi^2}$. To make sense of the equation
\EQ{fpsieq} we have to assume that $f$ is holomorphic and univalent in some
domain which includes the unit disk. Both sides of the equation have to match
in some annular region around the unit circle where both are simultaneously
meaningful. Alternatively, one can demand that both sides agree as formal power
series in the scaling parameter $z$, not to be confused with the coordinate
$\xi$. This is a well known problem in mathematics related to uniformization
and the existence of the Neretin semigroup \cite{Neretin1,Neretin2}.

Although in general it is more convenient to carry out computations in a
CFT-independent way, for this particular problem it is useful
to pick the simplest CFT corresponding to strings propagating freely in flat
space. The identity string field has a very simple expression and its
correlators can be easily evaluated by oscillator methods, see e.g.
\cite{GJ1,GJ2,KP}. Consider the following correlator
\be\label{eq:modelcorr}
\left(i\sqrt{2/\alpha'}\right)^2 \aver{ I \circ \partial X (x) U_1
z^{2L_0} U_1^*
\partial X(y)}.
\ee
Here we assume the total central charge is zero, so an insertion of
a weight zero operator like $c\partial c\partial^2 c$ is implicit.
We can compute the correlator in two different ways: Either
using formula \EQ{1psieq}, upon which we find the correlator is equal to
\be
\frac{\psi'(x)\psi'(y)}{(1+\psi(x)\psi(y))^2},
\ee
or we can compute it with the oscillator formalism. Let us commute $\partial X$
towards the center of the correlator and write it in its mode expansion
\be
i \sqrt{2/\alpha'}\, \partial X(w) = \sum_{n=-\infty}^\infty \alpha_n w^{-n-1}.
\ee
Next let us introduce normalized oscillators $a_n = \alpha_n /\sqrt{n}$ for $n
> 0$ and rewrite
\be
U_1^* \ket{0} = e^{-\frac{1}{2} \sum_{n=1}^\infty (-1)^n a_n^\dagger
a_n^\dagger} \ket{0}.
\ee
Using the formula
\be\label{eq:KPformula}
\aver{0| e^{\frac{1}{2} a.S.a} a_n a_m^\dagger e^{\frac{1}{2}
a^\dagger.V.a^\dagger} |0} = \det(1- S.V)^{-1/2} (1-V.S)^{-1}_{nm},
\ee
we find
\be
\frac{\psi'(x)\psi'(y)}{(1+\psi(x)\psi(y))^2} = \sum_{n=1}^\infty n z^{2n}
\left(\tilde x^{-n} + (-)^{n+1} \tilde x^{n}\right)\left(\tilde y^{-n}
+(-)^{n+1} \tilde y^{n}\right) \frac{1}{1-z^{4n}} \frac{1}{\tilde x \tilde y}
\frac{d\tilde x}{dx} \frac{d\tilde y}{dy},
\ee
where
\bea
\tilde x &=& x-\sqrt{1+x^2}, \nonumber\\
\tilde y &=& y+\sqrt{1+y^2}.
\eea
Note that thanks to the vanishing total central charge the determinant factor
from \EQ{KPformula} cancels against normalization constants from the
other sectors.

Imposing $\psi(0)=0$ the equation can be easily integrated. Expanding
$1/(1-z^{4n})$ into a geometric series the two infinite sums can be
interchanged and one finds
\be
1+\psi(x)\psi(y) = \prod_{k=0}^\infty \frac{(1-\frac{\tilde y}{\tilde x}
z^{4k+2})(1-\frac{\tilde x}{\tilde y} z^{4k+2})(1+\frac{1}{\tilde x \tilde y}
z^{4k+2})(1+\tilde x \tilde y z^{4k+2})}{(1-\frac{1}{\tilde x^2}
z^{4k+4})(1-\tilde x^2 z^{4k+4})(1-\frac{1}{\tilde y^2} z^{4k+4})(1-\tilde y^2
z^{4k+4})} (1-z^{8k+4})^2.
\ee
This equation at first sight seems rather unlikely to be self-consistent, the
right hand side does not look anything like one plus something factorizable.
Fortunately, the infinite product can be expressed in terms of Jacobi theta
functions\footnote{We use the notation of Polchinski, String Theory, Vol I.}:
\be
1+\psi(x)\psi(y) = \theta_4(1) \theta_3(1) \frac{\theta_4\left( \frac{\tilde
x}{\tilde y}\right) \theta_3\left(\tilde x \tilde y \right)}{\theta_4(\tilde
x)\theta_3(\tilde x)\theta_4(\tilde y)\theta_3(\tilde y)}.
\ee
The theta functions all depend on common {\it nome} $q=e^{2\pi i\tau}$ which we
suppressed and which is related to our previous scaling parameter $z$ by
$q=z^4$. Explicitly the theta functions are given by
\bea
\theta_3(x) &=& \sum_{n=-\infty}^\infty q^{n^2/2} x^n = \prod_{m=1}^\infty
(1-q^m) (1+x q^{m-1/2})(1+x^{-1} q^{m-1/2}), \\
\theta_4(x) &=& \sum_{n=-\infty}^\infty (-1)^n q^{n^2/2} x^n =
\prod_{m=1}^\infty
(1-q^m) (1-x q^{m-1/2})(1-x^{-1} q^{m-1/2}), \\
\theta_2(x) &=& \sum_{n=-\infty}^\infty q^{(n-1/2)^2/2} x^{n-1/2}
\nonumber\\
 &=& q^{1/8}(x^{1/2}+x^{-1/2}) \prod_{m=1}^\infty
(1-q^m) (1+x q^{m})(1+x^{-1} q^{m}), \\
\theta_1(x) &=& i \sum_{n=-\infty}^\infty (-1)^n q^{(n-1/2)^2/2} x^{n-1/2}
\nonumber\\
 &=& -i q^{1/8}(x^{1/2}-x^{-1/2}) \prod_{m=1}^\infty (1-q^m) (1-x
q^{m})(1-x^{-1} q^{m}).
\eea
From the representation in terms of infinite sums, one can easily derive an
identity
\be
\theta_4\left( \frac{\tilde x}{\tilde y}\right) \theta_3\left(\tilde x \tilde y
\right) = \frac{\theta_4(\tilde x)\theta_3(\tilde x)\theta_4(\tilde
y)\theta_3(\tilde y)}{\theta_4(1) \theta_3(1)} - \frac{\theta_1(\tilde
x)\theta_2(\tilde x)\theta_1(\tilde y)\theta_2(\tilde y)}{\theta_4(1)
\theta_3(1)}.
\ee
Using this identity the expression for $1+\psi(x)\psi(y)$ simplifies and we
find
\be 1+\psi(x)\psi(y) = 1-  \frac{\theta_1(\tilde x)\theta_2(\tilde
x)}{\theta_3(\tilde x)\theta_4(\tilde x)} \frac{\theta_1(\tilde
y)\theta_2(\tilde y)}{\theta_3(\tilde y)\theta_4(\tilde y)},
\ee
and hence
\be
\psi(x) = i \frac{\theta_1(\tilde x)\theta_2(\tilde x)}{\theta_3(\tilde
x)\theta_4(\tilde x)}  = q^{\frac{1}{4}}(\tilde x - \tilde x^{-1})
\prod_{m=1}^\infty \frac{1-\tilde x^2 q^{2m}}{1-\tilde x^2 q^{2m-1}}\,
\frac{1-\tilde x^{-2} q^{2m}}{1-\tilde x^{-2} q^{2m-1}}.
\ee
We see that indeed $\psi(0)=0$ and
\be
\psi'(0) = 2 q^{1/4}  \prod_{m=1}^\infty \left(\frac{1- q^{2m}}{1-
q^{2m-1}}\right)^2 = \frac{\eta(2\tau)^4}{\eta(\tau)^2}.
\ee
Now we can very easily compute the correlator \EQ{idcorr}:
\begin{equation}F(z)=
-\frac{1}{\pi^2}\frac{\eta(\tau)^4}{\eta(2\tau)^8},\ \ \ \ \ z=e^{i\pi\tau/2}.
\label{eq:ToyE}
\end{equation}
This function is holomorphic inside the unit circle $|z|<1$, but every point on
the unit circle is an essential singularity and the function cannot be
analytically continued beyond the unit disk
%(see figure \ref{fig:ToyE}).
(see figure \ref{fig:Pade_poles}b for the distribution of poles and zeros of
its Pad\'{e} approximant).

We can gain some intuition into the origin of these singularities by looking at
figure \ref{fig:simple_strip}. For $z=1$, the $c$ insertions sit right on top
of each other, but for $z>1$ the picture does not appear to make
sense---formally, the $c\,$s should be separated by a worldsheet of
``negative'' length. This is quite analogous to the worldsheet interpretation
of inverse wedge states, which are responsible for the divergence of the
$\L$ level expansion. Therefore it is not surprising that $F(z)$ is undefined
for $|z|>1$. Note also that the $F(z)$ occurs in the lower limit of integration
when we evaluate $\tilde{E}(z)$. Therefore figure \ref{fig:simple_strip} for
$z>0$ gives a nice intuitive picture for why the $L_0$ level expansion of
our solution is divergent.

%\begin{figure}
%\begin{center}
%\resizebox{2in}{2in}{\includegraphics{ToyE.eps}}
%\end{center}
%\caption{\label{fig:ToyE} Absolute value of the Dedekind ratio \EQ{ToyE}.
%Note that $F(z)$ vanishes at $z=1$.}
%\end{figure}

Now that we have a closed form solution for the level expansion, we can
evaluate $F(1)=\Tr[cQc]$ and see what we get\footnote{To prove this limit we
use the formula $\eta(-1/\tau)=\sqrt{-i\tau}\eta(\tau)$ and
$\eta(\tau) \sim e^{i\pi\tau/12}$ for large and positive $\mathrm{Im}(\tau)$.
Note that because $F(z)$
has essential singularities on the unit circle, in taking the limit $z\to 1$
we should be careful to follow a contour that intersects the real axis at an
angle of less than $90^\circ$.}:
\begin{equation}\lbra cQc\rbra=
-\frac{1}{\pi^2}\lim_{z\to1^-}\frac{\eta(\tau)^4}{\eta(2\tau)^8}
=0.\end{equation}
We have checked that this result agrees with the Pad\'e resummation of the
series \EQ{Polya}. In fact, we get the same answer
when computing in the $\mathcal{L}_0$ level expansion:
\begin{equation}\langle z^{\mathcal{L}_0}c,z^{\mathcal{L}_0}Q_B c\rangle=
-\frac{4}{\pi^2}
\left(\frac{1-z}{z}\right)^2.
\end{equation}
Again this vanishes at $z=1$. Given that $cQc$ is an identity-like string
field, it may be surprising that $\lbra cQc\rbra$ appears to vanish
regardless of the regularization---and even holds in the $L_0$ level
expansion\footnote{The fact that this trace vanishes is related to the negative
conformal dimension of $c$. A generic regularization of this correlator in the
$KBc$ subalgebra is $\lbra cQc\rbra = \lim_{\alpha\to 0}\lbra c\Omega^{\alpha
r_1}Bc\Omega^{\alpha r_2} Qc\Omega^{\alpha r_3}\rbra$ for $r_1,r_2,r_3 \ge 0$,
or linear combinations thereof. This vanishes as $\alpha^2$ due to the net
negative scaling dimension $-2$ of the insertions. Note that this has nothing
to do with the fact that $cQc$ vanishes in the Fock space; $cQcK^3$ also
vanishes in the Fock space, but its trace would generically be divergent by
this argument.}. There are actually good formal arguments for believing this
result. To see why, suppose we consider the energy of a vacuum solution $\Phi$
in the $\mathcal{L}_0^-$ level expansion. The energy function would be
\begin{equation}
E(z)=\frac{1}{6}\langle z^{\frac{1}{2}\L^-}\Phi, z^{\frac{1}{2}\L^-}
Q_B\Phi\rangle.\end{equation}
Because $\L^-$ is a reparameterization generator, this function is actually
independent of $z$. Now expand $\Phi$ in a basis of $\L^-$
eigenstates:
\begin{equation}\Phi \propto c + \mathrm{higher\ levels...}.\end{equation}
We can compute the energy alternatively as
\begin{equation}E(z) = \sum_{n=-2}^\infty z^nE_n,\end{equation}
where $E_n$ is the contribution to the action of fields whole total
$\frac{1}{2}\L^-$ level adds up to $n$. But since the energy is independent
of $z$, only the contribution $E_0$ can be nonvanishing, and in particular
\begin{equation}E_{-2}\propto \lbra cQc\rbra = 0,\end{equation}
consistent with the prediction of the $L_0$ and $\L$ level expansions. It would
be interesting to test this formal argument by extending the above computations
to the other $E_n$.

\section{Discussion}

In this paper we have given a simple analytic solution for
tachyon condensation in open bosonic string field theory. The absence of a
regulator and phantom term makes the solution easier to work with than in
$\B$ gauge. Moreover, the physics is much easier to see, as it is almost
exclusively contained in the term:
\begin{equation}c\frac{1}{1+K},\label{eq:0mtach}\end{equation}
which is nothing more than the zero momentum tachyon, albeit expressed in an
unusual gauge (see appendix \ref{app:gauge_fix}). The second
term
\begin{equation}cKBc\frac{1}{1+K}\end{equation}
is BRST exact, and its only purpose is to make the tachyon \EQ{0mtach}
satisfy the equation of motion. Of course, this fits nicely with the
intuition that the condensation of the tachyon field is really what's
responsible for the physics of tachyon condensation.

A novel feature of our solution is that it involves a continuous
superposition of wedge states arbitrarily close to the identity. The fact that
it is a continuous superposition, and not, say, an isolated identity-like
piece, is crucial for the consistency of our solution.
Indeed, many identity-based solutions have been proposed in the
past, but for such solutions there is no
unambiguous analytic calculation of the action; the level expansion is
divergent and cannot be meaningfully resummed\footnote{Though identity
based solutions are singular, some still correctly capture some nontrivial
open string physics. See especially \cite{Id5}.}. Still, there are
certain types of calculations that would be problematic for our solution.
For example, both $b(1)|\Psi\rangle$ and $b(1)|\hat{\Psi}\rangle$ are
divergent because the $b$ ghost gets ``too close'' to the
$c$ insertions inside $\Psi,\hat{\Psi}$. We hope that such issues will not
limit the utility of our solution.

Since the beginning, one of the great mysteries of string field theory has been
the remarkable success of the level expansion. One byproduct of our analysis
has been a much more detailed picture of why the level expansion works, and
in particular how it may fail to converge. It is quite remarkable that we
were able to solve the $L_0$ level expansion exactly for the field $c$---it
would be very interesting to find analogous solutions for other states.
Ideas along these lines could prove important for constructing a solution for
the tachyon vacuum in Siegel gauge.

There are many questions related to the tachyon vacuum that have yet to be
understood: Giving an analytic construction of the tachyon potential,
understanding vacuum string field theory and multiple D-branes
\cite{VSFT,OkawaVSFT,OkawaVSFT2}, recovering
closed string physics around the tachyon vacuum, and finding an analytic
tachyon vacuum in superstring field theory
\cite{super_photon,Fuchs_super,Kroyter1,Kroyter2,Kroyter3,ES}. Perhaps this
solution could inspire new approaches to marginal deformations
\cite{RZOK,Schnabl2,Erler4,Okawa_super,Okawa_super2,photon,KO,KO2},
or help in the construction of lump solutions \cite{Singular_gauge}. We hope
that our work will be useful for studying these important issues.

\bigskip
\noindent{\bf Acknowledgments}
\bigskip

\noindent We would like to thank Nathan Berkovits, David Gross, Michael
Kiermaier, Michael Kroyter, Yuji Okawa, Leonardo Rastelli and Barton Zwiebach
for useful discussion. We thank also Ian Ellwood for comments on the
manuscript. Both authors acknowledge warm hospitality of KITP where significant
portion of this research was done during the program Fundamental Aspects of
String Theory. MS would also like to thank the Rice Family Fund for generous
contribution that allowed him to bring his family to Santa Barbara for the
duration of the KITP program. This research was supported in part by the
National Science Foundation under Grant No. PHY05-51164 and in part by the
EURYI grant EYI/07/E010 from EUROHORC and ESF.

\begin{appendix}

\section{Star Products and Cylinder Correlators}
\label{app:cylinder}

In this appendix we explain how to translate expressions given in
the text into conformal field theory correlation functions on the cylinder.
The basic starting point are string fields $\Phi$ which can be represented
as a correlation function on a semi-infinite vertical strip of worldsheet
in the complex plane, with some operator insertions placed inside. The bottom
edge of the strip lies on the real axis, and corresponds to the boundary of
the open string; the ``top'' of the strip is at $+i\infty$, and corresponds
to the open string midpoint. On the positive and negative vertical edges of
the strip we impose boundary conditions corresponding to the left and right
halves of the open string\footnote{Fixing these boundary conditions requires
a choice of parameterization of the string along the vertical edges. Different
parameterizations correspond to different choices of projector conformal
frames \cite{RZO}. In this paper we have been using the sliver conformal
frame, where the standard parameterization of the half string with
$\sigma\in[0,\frac{\pi}{2}]$ maps to the vertical height
$y=\frac{1}{\pi}\tanh^{-1}\sin\sigma\in[0,\infty]$ on the strip edge. If
we had used the butterfly frame, the edges would be parameterized
as $y=\frac{1}{4}\tan\sigma\in[0,\infty]$.}, respectively. Evaluating the
resulting correlator gives a representation of $\Phi$ as a Schroedinger
functional of a classical open string configuration, $\sim \Phi[x(\sigma)]$.

Perhaps there is a possibility for geometrical confusion here, since the
{\it left} half of the string lies on the {\it right} (positive) edge of
the strip in the complex plane. This is an artifact of our star product
convention, which adheres to
\cite{Schnabl,Erler1,SZ,Review}. To solve this problem,
\cite{Okawa} introduced a different convention for the
star product with the opposite identification of left and right. We keep the
old convention, but to avoid confusing pictures it is helpful to visualize
the complex plane so that the positive real axis increases towards the
left---that is, our complex plane is related to the old one by $z\to-z^*$.
Then the left half of the string lies on the left (positive) boundary of the
strip. Note that closed contours in our visualization move clockwise---so our
convention might be called the {\it left handed} picture for the star
product, whereas that of \cite{Okawa} is the {\it right handed}
picture.

\begin{figure}[t]
\begin{center}
\resizebox{4.5in}{2.8in}{\includegraphics{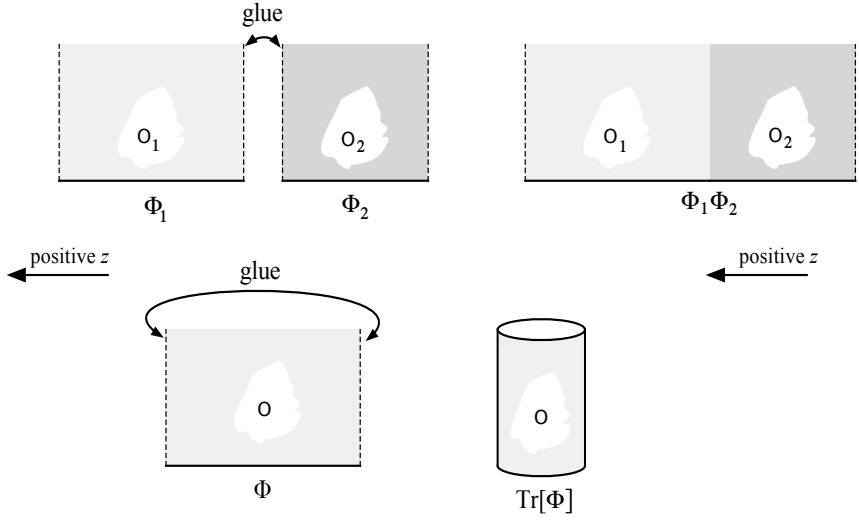}}
\end{center}
\caption{\label{fig:cyl} Star product and trace of open string functionals,
represented as correlation functions on a semi-infinite strip with possible
operator insertions. Note that if we visualize the real axis as increasing
towards the left, the order of the multiplication matches the geometrical
order of the gluing.}\end{figure}

Given a string field defined as a correlator on the strip, we can compute
star products and traces as follows: To compute the product
$\Phi_1\Phi_2[x(\sigma)]$, we glue $\Phi_1$'s negative
vertical edge to $\Phi_2$'s positive vertical edge, and evaluate the resulting
correlator. To compute the trace, we glue the positive and negative edges of
the strip together to form a correlation function on the cylinder. See figure
\ref{fig:cyl}. The gluing of edges is analogous to the contraction of
matrix indices---this is the essential intuition behind the split string
formalism \cite{Gross1,RSZ_half,Erler1}. Note that with our
picture of the complex plane, $\Phi_1$'s strip appears to the left of
$\Phi_2$'s in the product $\Phi_1\Phi_2[x(\sigma)]$, as would seem natural.

\begin{figure}[t]
\begin{center}
\resizebox{3.8in}{3.8in}{\includegraphics{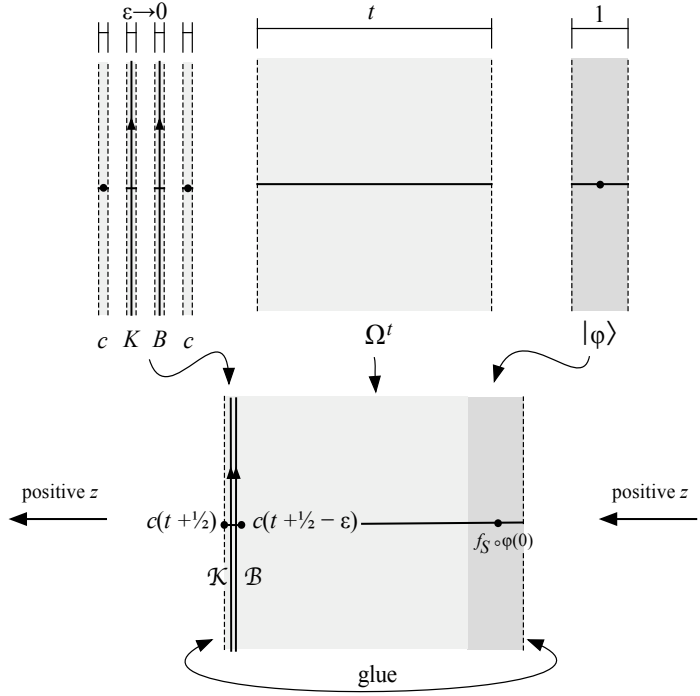}}
\end{center}
\caption{\label{fig:cyl2}Representation of the inner product $\lbra
cKBc\Omega^t \phi\rbra$ as a correlation function on the cylinder. The
parameter $\epsilon$ above is introduced for visual purposes, and should be
taken to zero. Note that positive $z$ increases from right to left in this
picture.}\end{figure}

Let us demonstrate how this works for fields in the $KBc$
subalgebra. We use the doubling trick to extend holomorphically to the lower \
half plane, so the semi-infinite vertical strip becomes an infinite vertical
strip extending from $-i\infty$ to $+i\infty$. The wedge state $\Omega^t$ is
then represented as an infinite vertical strip of worldsheet of width $t$,
without any operator insertions. A Fock space state
$|\phi\rangle = \phi(0)|0\rangle$ is a vertical strip of width 1, with an
insertion $f_\mathcal{S}\circ\phi(0)$ placed halfway between the edges of the
strip, on the real axis. Here
\begin{equation}f_\mathcal{S}(z) =\frac{2}{\pi}\tan^{-1}z \end{equation}
is called the sliver conformal map, and maps the unit disk to an
infinite vertical strip of width 1. Finally, consider the string fields
$K,B,c$. We take them to be {\it infinitely thin} vertical strips of
worldsheet carrying operator insertions
\begin{eqnarray}K\lineup\ \ \rightarrow\ \
\mathfrak{K}\equiv\int_{-i\infty}^{i\infty}\frac{dz}{2\pi i}T(z),\nonumber\\
B\lineup\ \ \rightarrow\ \
\mathfrak{B}\equiv\int_{-i\infty}^{i\infty}\frac{dz}{2\pi i}b(z),\nonumber\\
c\lineup\ \ \rightarrow\ \ c(z),
\end{eqnarray}
where $c(z)$ is inserted exactly on the strip, on the real axis. We can now
compute star products and traces of fields in the $KBc$ subalgebra by gluing
strip edges, as described above. The procedure is illustrated for
an example $\lbra cKBc\Omega^t \phi\rbra$ in figure \ref{fig:cyl2}.

Using this basic procedure, we can calculate the overlap of our solution
\EQ{solution} with any Fock space state:
\begin{equation}\lbra \Psi\phi\rbra = \int_0^\infty dt\, e^{-t}
\left\langle \left[c(t+\half) +
c(t+\half)\,\mathfrak{K}\mathfrak{B}\,
\lim_{\epsilon\to 0}c(t+\half-\epsilon)\right]f_\mathcal{S}\circ\phi(0)
\right\rangle_{C_{t+1}},\label{eq:CFTsol1}
\end{equation}
where $\langle\cdot\rangle_{C_{t+1}}$ is the correlation function on the
cylinder of circumference $t+1$ and the $\mathfrak{B}$ and $\mathfrak{K}$
contour insertions must be integrated between the $c$ ghosts on either side.
It is often convenient to represent the $\mathfrak{K}$ insertion as a
derivative of a wedge state $K = \left.\frac{d}{ds}\Omega^s\right|_{s=0}$.
Therefore we can also write
\begin{eqnarray}\lbra \Psi\phi\rbra = \lineup \int_0^\infty dt\, e^{-t}
\left[\Big\langle c(t+\half) f_\mathcal{S}\circ\phi(0)\Big\rangle_{C_{t+1}}
\right.\nonumber\\
\lineup\ \ \ \ \ \ \ \ \ \ \ \ \ \ \ \ \ \ \ \ \ \ \
+\left.\left.\frac{d}{ds}\Big\langle c(t+s+\half)\mathfrak{B}c(t+\half)
f_\mathcal{S}\circ\phi(0)\Big\rangle_{C_{t+s+1}}\right|_{s=0}\right] .
\label{eq:CFTsol2}\end{eqnarray}
Note that the gluing prescription does not determine the absolute location of
the operator insertions in the complex plane---it only determines their
relative positions, modulo the circumference of the cylinder. Here we have made
some convenient choice for the coordinates of the insertions.

Since both left and right handed star products have become common
in the literature, let us explain how to relate theories which use
these conventions. The right handed star product is related to the left
handed one by
\begin{equation}[AB]_R = (-1)^{AB}BA, \end{equation}
where the bracket $[\cdot]_R$ indicates that all star products inside are
right handed. We define a string field $A$ in our theory to be
{\it equivalent} to a string field $A'$ in the right handed theory if they
are related by:
\begin{equation}A' =  A^\S, \label{eq:LRmap}\end{equation}
where $A^\S = (-1)^{L_0}A$ denotes twist conjugation, a graded involution of
the star product corresponding to a reversal of the parameterization of the
open string\footnote{$A^\S$ is related to the twist conjugation introduced in
\cite{Zwiebach,Review} by a minus sign. Thus a twist even solution acquires a
minus sign under conjugation with $\S$.} \cite{Zwiebach,Review}. This
involution satisfies
\begin{equation}(QA)^\S = Q(A^\S),\ \ \ \ \ (AB)^\S = (-1)^{AB}B^\S A^\S,\ \
\ \ \ \Llbra A^\S\Rrbra = \lbra A\rbra.\end{equation}
For fields in the $KBc$ subalgebra
\begin{equation}c^\S = -c,\ \ \ \ \ K^\S = K,\ \ \ \ \ B^\S = B.\end{equation}
If string fields in the left and right handed theory are related by this
twist, one can show:
\begin{equation}[f(A',B',...)]_R = f(A,B,...)^\S,\end{equation}
where $f$ is any function of a list of string fields. This has two
consequences: First, if we have a relation between
string fields of the form
\begin{equation}f(A,B,...)=0,\end{equation}
then the corresponding relation holds in the right handed theory:
\begin{equation}[f(A',B',...)]_R = 0.\end{equation}
Second, traces between the two theories agree:
\begin{equation}\Blbra [f(A',B',...)]_R\Brbra =
\Blbra \,f(A,B,...)\, \Brbra.\end{equation}
Therefore we know how to translate any statement about string fields in our
left handed convention to a statement about string fields in the right handed
convention. One can check that the $\B$ gauge vacuum picks up an extra sign
under twist conjugation, which accounts for the sign discrepancy between
the solutions presented in \cite{Schnabl} and \cite{Okawa}. Our
solution $\Psi$ maps to
\begin{equation}\Psi' = \frac{1}{1+K}(-c+cKBc) =
-\left[(c+cKBc)\frac{1}{1+K}\right]_R.\end{equation}
Note that in the right handed convention, the sign in front of $c$ insertion
is negative. This is because in the right handed picture the tachyon condenses
towards the left of the perturbative vacuum in the tachyon potential.

\section{Equivalence to the $\B$ Gauge Solution}
\label{app:bgauge}

In this appendix we explicitly construct the gauge parameter relating our
solution to the $\B$ gauge solution. Consider two dressed $\B$ gauge
solutions
\begin{equation}\Phi = fc\frac{KB}{1-fg}cg,\ \ \ \ \ \ \
\Phi' = f'c\frac{KB}{1-f'g'}cg',\end{equation}
where $f,f',g,g'$ are functions of $K$. If these solutions are gauge
equivalent, they can be related by the transformation
\begin{equation}\Phi' = U^{-1}(Q+\Phi)U,\end{equation}
where
\begin{eqnarray}U = \lineup 1-fBc\,g+\left(\frac{1-fg}{1-f'g'}\right)
f'Bc\,g',\nonumber\\
U^{-1} = \lineup 1-f'Bc\,g'+\left(\frac{1-f'g'}{1-fg}\right)
fBc\,g.\end{eqnarray}
If they are not gauge equivalent, than either $U$ or $U^{-1}$ must be
singular. The only part of the above expressions which could
potentially cause problems are the factors in parentheses.
Therefore, $\Phi$ and $\Phi'$ are gauge equivalent if and only if the string
field
\begin{equation}M = \frac{1-fg}{1-f'g'}\end{equation}
and its inverse are well defined. In practice, the easiest way to see this
is to check that both $M$ and $M^{-1}$ are analytic functions of $K$ at $K=0$
\footnote{We do not have a complete understanding of what constitutes an
acceptable state in the wedge algebra. It seems necessary that the state is a
$C^\infty$ function of $K$ at $K=0$, but we further assume that it should be
analytic. Still this condition is not sufficient. Though $M$ and $M^{-1}$ 
may be
analytic at $K=0$, they may not be expressible in terms of non-negative powers
of the $SL(2,\mathbb{R})$ vacuum. But if this is the case, either $\Phi$ or
$\Phi'$ would be a string field built out of inverse wedge states.
Therefore, if we assume $\Phi,\Phi'$ are well behaved, the power series
argument is sufficient to demonstrate their gauge equivalence.}.
Since $fg$ and $f'g'$ must also be
analytic, this amounts to the requirement that the first nonvanishing powers
in a Taylor series expansion of $1-fg$ and $1-f'g'$ must be the same:
\begin{equation}1-fg \sim K^n + \mathrm{higher\ powers...},\ \ \ \ \ \
1-f'g' \sim K^n
+\mathrm{higher\ powers...}.\end{equation}
The integer $n$ plays the role of an index labeling physically inequivalent
solutions in the $KBc$ subalgebra. $n=0$ describes the perturbative vacuum and
$n=1$ describes the closed string vacuum. Other possible values of $n$ are
mysterious since the corresponding solutions do not appear to be well-defined.
They have been conjectured to be related to multiple brane solutions
\cite{multiple}.

For the $\B$ gauge vacuum and our new solution, we have
\begin{eqnarray}1-fg\ = \lineup\, \frac{K}{1+K} = K + \mathrm{higher\
powers...},
\nonumber\\
1-f'g'= \lineup\ 1-\Omega\  = K+\mathrm{higher\ powers...}.\end{eqnarray}
Therefore the solutions are gauge equivalent and describe the closed string
vacuum. Explicitly, $M$ and $M^{-1}$ are,
\begin{eqnarray}M = \lineup \lim_{N\to\infty}
\int_0^\infty dt e^{-t}\left[\Omega^{N+t}-\sum_{n=0}^N \frac{d}{dt}
\Omega^{n+t}\right],\nonumber\\
M^{-1} = \lineup 1-\Omega +\int_0^1 dt\, \Omega^t.\end{eqnarray}
Note the presence of a limit and sliver-like term in the expression for $M$.
This is the origin of the regulator and phantom piece in the $\B$ gauge
solution.

\section{Gauge fixing}
\label{app:gauge_fix}

In this appendix we give a setup for understanding the gauge fixing of the
new solution (\ref{eq:solution}, \ref{eq:solution2}) and related solutions
appearing in \cite{Erler2}. To this end, we define the operator
\begin{equation}\mathcal{B}_{f,g}\Phi =
\frac{1}{2}f[\B^-(f^{-1}\Phi g^{-1})]g,\end{equation}
where $f,g$ are functions of $K$ and $\B^-=\B-\B^*$. Also define
\begin{equation}\mathcal{L}_{f,g}\Phi =
\frac{1}{2}f[\L^-(f^{-1}\Phi g^{-1})]g.\end{equation}
These operators are easy to evaluate on wedge states with
insertions since $\B^-,\L^-$ are derivations and
\begin{eqnarray}
\frac{1}{2}\B^- K =\lineup B,\ \ \ \ \ \ \
\frac{1}{2}\L^- K = K,\nonumber\\
\frac{1}{2}\B^-B = \lineup 0,\ \ \ \ \ \ \ \ \
\frac{1}{2}\L^- B = B,\nonumber\\
\frac{1}{2}\B^- c = \lineup 0,\ \ \ \ \ \ \ \ \ \
\frac{1}{2}\L^- c = -c.\end{eqnarray}
We should think of $\mathcal{B}_{f,g},\mathcal{L}_{f,g}$ as generalizations
of $\B,\mathcal{L}_0$. In fact
\begin{equation}\mathcal{L}_{F,F}=\mathcal{L}_0,\ \ \ \
\mathcal{B}_{F,F}=\mathcal{B}_0,\end{equation}
where $F=\sqrt{\Omega}$ is the square root of the $SL(2,\mathbb{R})$ vacuum.
In particular, $\B$ gauge is just an example of a
large family of gauges
\begin{equation}\mathcal{B}_{f,g} \Phi =0 .\end{equation}
Note that the string field must be ``dressed'' by factors of
$f^{-1},g^{-1}$ before it is annihilated by $\B^-$. For this reason, we
call these {\it dressed} $\B$ gauges. The new solutions
$\Psi$ and the real $\hat\Psi$ satisfy gauge
conditions of this type:
\begin{eqnarray}\mathcal{B}_{1,\frac{1}{1+K}}\Psi =\lineup 0,\label{eq:ass_g}\\
\mathcal{B}_{\frac{1}{\sqrt{1+K}},\frac{1}{\sqrt{1+K}}}\hat{\Psi}=
\lineup 0. \end{eqnarray}
Equation (\ref{eq:ass_g}) can be reexpressed in a particularly simple form:
\begin{equation}\B^-\left(1-\frac{\pi}{2}(K_1)_R\right)\Phi=0.\end{equation}
It could be interesting to explore the consequences of these gauges in
perturbation theory.

Of all these gauges, $\B$ gauge certainly appears to be the
most natural one. It is reasonable to wonder, then, in what sense our new gauge
$\mathcal{B}_{\frac{1}{\sqrt{1+K}},\frac{1}{\sqrt{1+K}}}\Phi=0$ is
special or unique. One answer to this question is given by the level
expansion. Given any solution satisfying a linear gauge condition
$\mathcal{O}\Phi=0$, one can define a ``natural'' level expansion in
terms of eigenstates of the operator $[Q_B,\mathcal{O}]$.
For Siegel gauge, this leads to the ordinary $L_0$ level expansion;
for $\B$ gauge, this gives the $\mathcal{L}_0$ level expansion. For the
new solution $\hat{\Psi}$, the natural expansion is in terms of eigenstates of
$\mathcal{L}_{\frac{1}{\sqrt{1+K}},\frac{1}{\sqrt{1+K}}}$. Remarkably, this
expansion of \EQ{solution2} terminates after just two levels:
\begin{equation}
\mathrm{Level\,-1}:\ \ \ \frac{1}{\sqrt{1+K}}c\frac{1}{\sqrt{1+K}},\ \ \ \ \
\mathrm{Level\, 0}:\ \ \ \frac{1}{\sqrt{1+K}}cKBc\frac{1}{\sqrt{1+K}}.
\end{equation}
Indeed this is remarkable---certainly we do not find the tachyon condensate
in Siegel gauge after expanding out to level 2. In fact, this can be taken as
the defining property of our solution, according to the following claim:

\bigskip
\noindent {\bf Claim:} Eq.(\ref{eq:solution}) is the unique, regular
dressed $\B$ gauge solution in the $KBc$ subalgebra that terminates at
finite level in its own level expansion, up to homogeneous gauge
transformations.
\bigskip

\noindent We can establish this as follows. For a solution to
terminate at level
$n-1$ in its own level expansion, the function of $K$ sandwiched between the
$c$ insertions must be an $n$th degree polynomial, call it $P_n$. The non-real
form of the solution is then
\begin{equation}\Phi=c BP_n c\left(1-\frac{K}{P_n}\right),\label{eq:level_sol}
\ \ \ \ \ \ \ \ \mathcal{B}_{1,1-\frac{K}{P_n}}\Phi=0.
\end{equation}
It is helpful to cancel the $K$ in the numerator. Assuming $n\geq 1$, $P_n$
has at least one root, which we can call $-\frac{1}{\gamma}$. Then write
$P_n = \left(K+\frac{1}{\gamma}\right)\pi_{n-1}$ with $\pi_{n-1}$ some
polynomial of order $n-1$, and the solution becomes
\begin{equation}\Phi=c BP_n c\left(1-\frac{1}{\pi_{n-1}}+
\frac{1}{\gamma}\frac{1}{P_n}\right).
\label{eq:level_sol2}\end{equation}
The first term is the identity string field with some insertions. Unless
the identity piece cancels, the action evaluated on the solution will be
undefined\footnote{Note also that the trace of an identity-like string field
is undefined if the field carries insertions with total zero
or positive scaling dimension in the sliver coordinate frame. This is
certainly true of \EQ{level_sol2}.}. For $n\geq 2$, the inverses of $P_n$
and $\pi_{n-1}$ can be
found by making a partial fraction decomposition and expressing the resulting
terms as integrals over wedge states via the Schwinger parameterization.
None of this produces a piece which would cancel the identity string field,
so for $n\geq 2$ the solutions are ill-defined. However, for $n=1$,
$\pi_{n-1}=\pi_0$ is a constant; if we choose $\pi_0=1$ the identity is
exactly canceled, leaving $P_n = \frac{1}{\gamma}+K$ and
\begin{equation}\Phi = \left(\frac{1}{\gamma}c+cKBc\right)\frac{1}{1+\gamma K}.
\end{equation}
This is our original solution \EQ{solution}, up to a reparameterization
$\gamma^{\mathcal{L}_0^-/2}$. This leaves the case $n=0$; the solution
there is
\begin{equation}\Phi=\frac{1}{\gamma}c(1-\gamma K).\end{equation}
This is a singular identity-based solution. Therefore only $n=1$ admits
a regular solution to the equations of motion, as claimed.

Let us list a few useful properties of dressed $\B$ operators. Dressed $\B$
operators have the following symmetries under conjugation:
\begin{eqnarray}
\mathcal{B}_{f,g}^{\,*} = \lineup-\mathcal{B}_{f^{-1},g^{-1}},\\
\mathcal{B}_{f,g}^{\,\dag} =
\lineup -\mathcal{B}_{\bar{g}^{-1},\bar{f}^{-1}},\\
\mathcal{B}_{f,g}^{\,\ddag} = \lineup \mathcal{B}_{\bar{g},\bar{f}}
\label{eq:Bddag}\\
\mathcal{B}_{f,g}^{\,\S} = \lineup \mathcal{B}_{g,f}\label{eq:BS}.
\end{eqnarray}
Here, $*$ denotes BPZ conjugation, $\dag$ denotes Hermitian conjugation,
$\ddag$ is reality conjugation, $\S$ is twist conjugation, and
$\bar{f},\bar{g}$ are the complex conjugates of $f,g$. The same properties also
hold for $\mathcal{L}_{f,g}$. Note that equations
(\ref{eq:Bddag},\ref{eq:BS}) imply that a dressed
$\B$ gauge solution is consistent with the reality condition only when
$f=\bar{g}$, and it is twist even only when $f=g$.

To give some other formulas, it is helpful to introduce the string fields,
\begin{equation}B_f = B f\frac{d}{dK}f^{-1},\ \ \ \ \
K_f = K f\frac{d}{dK}f^{-1}.\end{equation}
We have for example,
\begin{equation}
B_1=0,\ \ \ \ \ \ B_\Omega=B,\ \ \ \ \ \
B_{\frac{1}{1+K}}=\frac{B}{1+K}.\end{equation}
$B_f$ and $K_f$ characterize the failure of
$\mathcal{B}_{f,g},\mathcal{L}_{f,g}$ to be derivations of the star product:
\begin{eqnarray}
\mathcal{B}_{f,g}(\Phi\Lambda) =
\lineup\left(\mathcal{B}_{f,v}\Phi\right)\Lambda +
(-1)^\Phi \Phi\left(\mathcal{B}_{u,g}\Lambda\right)
-(-1)^\Phi \Phi B_{uv}\Lambda,\\
\mathcal{L}_{f,g}(\Phi\Lambda) =
\lineup\left(\mathcal{L}_{f,v}\Phi\right)\Lambda +
\Phi\left(\mathcal{L}_{u,g}\Lambda\right)-\Phi K_{uv}\Lambda.
\end{eqnarray}
To give a slightly more general formula we have introduced arbitrary
$u,v$ on the right hand side. Note that this implies that
$\mathcal{B}_{f,f^{-1}},\mathcal{L}_{f,f^{-1}}$ are derivations of the star
product. Also note
\begin{equation} \mathcal{B}_{f,g}|I\rangle = B_{fg},\ \ \ \ \
\mathcal{L}_{f,g}|I\rangle = K_{fg}.\end{equation}
Two dressed $\B$ operators can be related by left/right
multiplication with $B_f$:
\begin{equation}\mathcal{B}_{f,g}\Phi = \mathcal{B}_{u,v}\Phi
+ B_{f/u}\Phi+(-1)^\Phi \Phi B_{g/v}\end{equation}
with a similar formula for $\mathcal{L}_{f,g}$. $B_f$ and $K_f$ satisfy a
logarithmic sum/product rule:
\begin{equation}aB_f+bB_g =  B_{f^ag^b},\ \ \ \ \ a,b\in\mathbb{C}
\end{equation}
which implies a similar rule for $\mathcal{B}_{f,g},\mathcal{L}_{f,g}$:
\begin{equation}
a\mathcal{B}_{f,g} +b\mathcal{B}_{h,j} =
\mathcal{B}_{f^ah^b,g^aj^b},\ \ \ \ \ a,b\in\mathbb{C},\ \ a+b=1.\end{equation}
The restriction $a+b=1$ gives a simpler formula, but the general case follows
by multiplying this equation by a constant. Thus dressed $\B,\mathcal{L}_0$
operators form a closed linear space; in particular, we cannot make new
gauges by taking linear combinations of $\mathcal{B}_{f,g}$s.

The special projector algebra \cite{Schnabl,RZ}
$[\mathcal{L}_0,\mathcal{L}_0^*]=\mathcal{L}_0+\mathcal{L}_0^*$ plays an
important role in the algebraic structure of analytic solutions. There is an
analogue of this algebra for dressed $\mathcal{L}_0$ operators. To display
this algebra is is useful to introduce a ``dressed'' analogue of a wedge
state:
\begin{equation}\Omega(f) = e^{-K_f},\end{equation}
and,
\begin{equation}\Omega(f^ag^b) = \Omega(f)^a\Omega(g)^b\ \ \ a,b\in\mathbb{C}.
\end{equation}
The generalization of the special projector algebra is then,
\begin{equation}[\mathcal{L}_{f,g},\mathcal{L}_{u,v}^*] =
\mathcal{L}_{\Omega(f),\Omega(g)}+\mathcal{L}_{\Omega(u),\Omega(v)}^*
\end{equation}
Note that $\Omega(\cdot)$ acts as the identity on wedge states, so we recover
the usual formula when $f=u=F$ and $g=v=F$.

\end{appendix}

\end{document}